\documentclass[journal]{IEEEtran}
\pdfoutput=1
\IEEEoverridecommandlockouts

\usepackage{acronym}
\usepackage{amsmath,amssymb}
\usepackage[english]{babel}
\usepackage{bbm}
\usepackage{bm}
\usepackage{caption}
\usepackage{cite}
\usepackage{enumerate}
\usepackage[T1]{fontenc}
\usepackage{graphicx}
\usepackage{hyperref}
\usepackage{cleveref}
\usepackage[utf8]{inputenc}
\usepackage{siunitx}
\usepackage{subcaption}
\usepackage{xcolor}

\usepackage{tikz}
\usetikzlibrary{math}
\usetikzlibrary{shapes.geometric,calc}
\usetikzlibrary{arrows.meta, quotes, angles}
\usetikzlibrary{decorations}
\usetikzlibrary{chains,shapes.misc,positioning,scopes}
\usepackage{circuitikz}
\usetikzlibrary{circuits.ee.IEC}
\usetikzlibrary{decorations}
\usetikzlibrary{dsp}
\usetikzlibrary{shadows}

\definecolor{myblue}{RGB}{31,119,180}
\definecolor{myorange}{RGB}{255,127,14}
\definecolor{mygreen}{RGB}{44,160,44}

\usepackage{nicefrac}

\acrodef{DMC}{Diffusive Molecular Communication}
\acrodef{ISI}{inter-symbol interference}
\acrodef{MC}{Molecular Communication}
\acrodef{NT}{neurotransmitter}
\acrodef{wrt}[w.r.t.]{with respect to}
\acrodef{wlog}[w.l.o.g.]{without loss of generality}
\acrodef{TFM}{transfer function model}
\acrodef{SSD}{state-space description}
\acrodef{PBS}{particle-based simulations}
\acrodef{CSK}{concentration-shift keying}
\acrodef{ML}{maximum likelihood}
\acrodef{CSI}{channel-state information}
\acrodef{BER}{bit-error rate}
\acrodef{EPSP}{excitatory postsynaptic potential}
\acrodef{iid}[i.i.d.]{independent and identically distributed}

\newcommand{\dtau}{\ensuremath{\mathrm{d}\tau}}
\newcommand{\ka}{\ensuremath{\kappa_a}}
\newcommand{\kd}{\ensuremath{\kappa_d}}
\newcommand{\kad}{\ensuremath{\kappa_{a_0}}}

\newcommand{\iinfty}{\ensuremath{i^{\infty}}}
\newcommand{\xmin}{\ensuremath{x_{\mathrm{min}}}}
\newcommand{\xmax}{\ensuremath{x_{\mathrm{max}}}}
\newcommand{\Omegax}{\ensuremath{\Omega_{\textrm{x}}}}
\newcommand{\kece}{\ensuremath{\kappa_e C_E}}

\newcommand{\tmax}{\ensuremath{t_{\mathrm{max}}}}

\newcommand{\dint}[1]{\,\mathrm{d}#1}
\newcommand{\Kprim}{\bm{K}}
\newcommand{\Kadj}{\tilde{\bm{K}}}

\newcommand{\tran}{^{\scriptscriptstyle\mathrm{T}}}
\newcommand{\Fbh}{\bm{F}_\mathrm{b}\tran}
\newcommand{\As}{\bm{\mathcal{A}}}
\newcommand{\Cs}{\bm{\mathcal{C}}}
\newcommand{\Csa}{\tilde{\bm{\mathcal{C}}}}
\newcommand{\Ks}{\tilde{\bm{\mathcal{K}}}}

\newtheorem{theorem}{Theorem}
\newtheorem{lemma}{Lemma}

\newtheorem{corollary}{Corollary}
\newtheorem{remark}{Remark}

\allowdisplaybreaks

\makeatletter
\long\def\@makecaption#1#2{\ifx\@captype\@IEEEtablestring%
    \footnotesize\begin{center}{\normalfont\footnotesize #1}\\
        {\normalfont\footnotesize\scshape #2}\end{center}%
    \@IEEEtablecaptionsepspace
    \else
    \@IEEEfigurecaptionsepspace
    \setbox\@tempboxa\hbox{\normalfont\footnotesize {#1.}~~ #2}%
    \ifdim \wd\@tempboxa >\hsize%
    \setbox\@tempboxa\hbox{\normalfont\footnotesize {#1.}~~ }%
    \parbox[t]{\hsize}{\normalfont\footnotesize \noindent\unhbox\@tempboxa#2}%
    \else
    \hbox to\hsize{\normalfont\footnotesize\hfil\box\@tempboxa\hfil}\fi\fi}
\makeatother

\begin{document}
    
\title{Saturating Receiver and Receptor Competition\\ in Synaptic DMC:\\ Deterministic and Statistical Signal Models}
\author{\IEEEauthorblockN{Sebastian Lotter, Maximilian Sch\"afer, Johannes Zeitler, and Robert Schober}\\
    \IEEEauthorblockA{\small Friedrich-Alexander University Erlangen-Nuremberg, Germany} \thanks{This paper has been accepted for presentation in part at the IEEE International Conference on Communications (ICC), 2021. This work was supported in part by the German Research Foundation (DFG) under grant SCHO 831/9-1.}
}

\maketitle
\nocite{lotter20b}

\begin{abstract}
Synaptic communication is based on a biological \ac{MC} system which may serve as a blueprint for the design of synthetic \ac{MC} systems.
However, the physical modeling of synaptic \ac{MC} is complicated by the possible saturation of the molecular receiver caused by the competition of \acp{NT} for postsynaptic receptors.
Receiver saturation renders the system behavior nonlinear in the number of released \acp{NT} and is commonly neglected in existing analytical models.
Furthermore, due to the ligands' competition for receptors (and vice versa), the individual binding events at the molecular receiver are in general not statistically independent and the commonly used binomial model for the statistics of the received signal does not apply.
Hence, in this work, we propose a novel deterministic model for receptor saturation in terms of a state-space description based on an eigenfunction expansion of Fick's diffusion equation.
The presented solution is numerically stable and computationally efficient.
Employing the proposed deterministic model, we show that saturation at the molecular receiver effectively reduces the peak-value of the expected received signal and accelerates the clearance of \acp{NT} as compared to the case when receptor occupancy is neglected.
We further derive a statistical model for the received signal in terms of the hypergeometric distribution which accounts for the competition of \acp{NT} for receptors {\em and} the competition of receptors for \acp{NT}.
The proposed statistical model reveals how the signal statistics are shaped by the number of released \acp{NT}, the number of receptors, and the binding kinetics of the receptors, respectively, in the presence of competition.
In particular, we show that the impact of these parameters on the signal variance is qualitatively different depending on the relative numbers of \acp{NT} and receptors.
Finally, the accuracy of the proposed deterministic and statistical models is verified by particle-based computer simulations.
\end{abstract}

\section{Introduction}
\acresetall
\ac{MC} is a bio-inspired communication paradigm in which information is transmitted via molecules.
It has gained significant attention as potential enabler of novel applications in the context of the Internet of Bio-nano Things \cite{akyildiz15}.
In particular, \ac{MC} is considered as a promising candidate for novel intra-body applications due to its inherent bio-compatibility and the fact that traditional electromagnetic wave-based wireless communication is not feasible at nano-scale \cite{nakano13}.
Natural \ac{MC} systems have evolved over millions of years to cope with the challenges faced in intra-body nano-scale communication and might hence serve as blueprints for synthetic \ac{MC} systems.
Among the different natural types of \ac{MC}, \ac{DMC}, i.e., communication via diffusing molecules, is a promising candidate for synthetic \ac{MC} as it requires neither dedicated communication infrastructure nor external energy sources for molecule propagation \cite{nakano13}.

\acp{DMC} can be found in the human body for example in chemical synapses formed between adjacent nerve cells or between neurons and muscle fibers.
The synaptic \ac{MC} system comprises in its simplest form the presynaptic cell (transmitter), the postsynaptic cell (receiver), and the synaptic cleft (channel) \cite{hof14}, cf.~Fig.~\ref{fig:channel}.
The message carrying molecules are termed \acp{NT}.
To convey information, \acp{NT} are released from vesicular containers at the presynaptic cell, diffuse across the synaptic cleft, and bind to postsynaptic receptors \cite{nakano13}.
The synaptic signaling is terminated as \acp{NT} are removed from the synaptic space by uptake or enzymatic degradation \cite{zucker14}.

In the synaptic \ac{DMC} system, the molecular receiver comprises a finite number of postsynaptic receptors for which the \acp{NT} compete.
This competition results in a phenomenon called {\em receptor saturation} encountered at some common types of synapses in the mammalian brain \cite{foster05}.
The term ``receptor saturation'' is sometimes used in the \ac{MC} literature for the case that (almost) {\em all} available receptors are occupied \cite{kuscu18}.
Here, however, we adopt a more general view and refer to a receiver that comprises a finite number of individual receptors to each of which a finite number of molecules may bind reversibly as {\em saturating receiver}.
The saturating receiver is a special type of {\em reactive receiver} \cite{jamali19}.
It reduces to a {\em reversibly absorbing receiver} if the number of receptors approaches infinity or the molecules remain bound to the receptors only very briefly \cite{ahmadzadeh16}.

The impact of saturation on the received signal at the molecular receiver is twofold.
First, it limits the maximum number of molecules concurrently bound to receptors.
Hence, the {\em expected received signal} at the saturating receiver is {\em nonlinear} in the number of released molecules.
Second, in contrast to linear receivers, the {\em signal statistics} at the saturating receiver do not follow the binomial distribution (which is commonly used in the \ac{MC} literature to model the received signal statistics \cite{jamali19}), because the individual binding events are {\em statistically dependent}.
Consequently, saturation renders the analytical characterization of the saturating receiver in terms of both the mean received signal and the signal statistics a challenging problem.

Synaptic communication has been studied in the \ac{MC} literature before and we refer the reader to \cite{veletic2019,lotter20} for recent literature overviews.
In most models for the synaptic communication channel, e.g.~in \cite{balevi13,liu14,veletic16b,lotter20a}, postsynaptic receptor saturation is neglected.
In \cite{khan2017}, a finite number of postsynaptic receptors are assumed, but the impact of \ac{NT} buffering at the postsynaptic receptors on the concentration of solute \acp{NT} is not taken into account. 
This approach leads potentially to an underestimation of the synaptic \ac{ISI} caused by residual \acp{NT} in the synaptic cleft.
Receptor saturation and its impact on the concentration of solute \acp{NT} in the presence of presynaptic \ac{NT} transporters is modeled in \cite{bilgin17}.
The resulting nonlinear model is solved by discretizing the diffusion equation in space and time and employing an iterative numerical algorithm.
In \cite{veletic19a}, a nonlinear model for ligand-receptor binding based on a system of ordinary differential equations is studied using Volterra series.
The spatial distribution of molecules is, however, not considered in \cite{veletic19a}.

Saturation at the molecular receiver has been considered in the \ac{MC} literature also in the context of targeted drug delivery \cite{femminella15,felicetti16,salehi19} and experimental studies \cite{farsad14,kim15,kim19}.
In none of these works, however, the impact of receptor saturation on the spatial distribution of solute molecules is modeled explicitly.

In \cite{ahmadzadeh16}, a deterministic model for a reversibly binding reactive receiver is proposed and receptor saturation is studied in an unbounded environment using \ac{PBS}.
In \cite{sun20}, the deterministic model from \cite{ahmadzadeh16} is extended to incorporate the impact of finitely many receptors on the reactive receiver in an unbounded environment.
The iterative scheme presented in \cite{sun20}, however, is computationally very expensive and only approximate when compared to \ac{PBS}.

The statistics of a molecular receiver comprising finitely many receptors are studied in \cite{pierobon11a}.
However, the model presented in \cite{pierobon11a} is limited to bandlimited input signals which are independent of the molecular binding events at the receiver.
A statistical signal model for the reactive receiver without receptor saturation was presented in \cite{liu20}.
In \cite{liu20}, however, the saturation of the molecular receiver is neglected, i.e., molecules are assumed to bind statistically independently to receptors.
As the probability for a molecule to find a free receptor decreases when more receptors are occupied, this assumption is in general not satisfied.
In \cite{kuscu18}, a statistical model for the received signal in the presence of receptor saturation is proposed based on the binomial distribution assuming mutual statistical independence between the receptors.
However, the more molecules are bound to receptors, the less likely a receptor is hit by a solute molecule, i.e., receptors compete for molecules \cite{berezhkovskii12}.
Hence, the assumption of statistical independence of receptors is in general also not satisfied.

In this paper, we propose a novel deterministic and a novel statistical signal model for the saturating receiver in synaptic \ac{DMC} in the presence of enzymatic degradation.
The proposed deterministic model is based on the diffusion equation and incorporates an analytical model of the reversible binding of \acp{NT} to a finite number of postsynaptic receptors in terms of a saturation boundary condition.
In contrast to previous works, our deterministic model encompasses a spatial model of the synaptic cleft and a finite number of postsynaptic receptors without decoupling the concentrations of solute and bound molecules (as, e.g., in \cite{khan2017}) or the need for spatial discretization (as, e.g., in \cite{bilgin17}).
In contrast to the deterministic model proposed in \cite{sun20}, it matches the results obtained with \ac{PBS} very accurately.
Our approach is based on the modeling of the diffusion equation in terms of a \ac{SSD} \cite{schaefer:ecc:2019}.
It utilizes a functional transformation of the diffusion equation adapted to the synaptic geometry and allows the modular incorporation of the nonlinear receptor saturation effect by a feedback structure \cite{schaefer:laminar:2020}.
Compared to particle-based Monte Carlo methods, the approach presented in this paper is computationally extremely efficient as the computational cost scales neither with the number of released particles nor with the number of receptors.
Furthermore, it yields the expected received signal without diffusion noise.
To the best of the authors' knowledge, the proposed deterministic signal model is the first analytical model that simultaneously takes into account enzymatic degradation and receptor saturation in a bounded domain.

Starting from the proposed deterministic model, we use steady state analysis to derive a novel statistical signal model for the saturating receiver in terms of the hypergeometric distribution.
The competition of molecules for receptors (receptors for molecules) causes negative correlation between the stochastic binding of individual molecules to receptors (receptors to molecules).
In contrast to existing statistical models, the proposed model captures this statistical dependence and thereby allows for more accurate modeling of the received signal statistics in the presence of competition.
In particular, we show that the existing binomial models tend to overestimate the variance of the received signal when compared to the proposed hypergeometric model.
Using the proposed statistical model, we examine the impact of the number of molecules, the number of receptors, and the binding rate of molecules to receptors on the variance of the received signal.
Our results suggest that, while increasing any of these parameters leads to an increase of the peak value of the expected received signal, each of these parameters impacts the signal statistics in a different manner.
The insights obtained by this analysis can be especially helpful for \ac{DMC} system design.
The excellent accuracy of our model is confirmed by \ac{PBS}.

The deterministic signal model presented in this paper was introduced in part in \cite{lotter20b}.
In contrast to \cite{lotter20b}, however, in the present paper, the deterministic signal model is complemented by a statistical signal model.
This extension facilitates the analysis of the joint impact of the system parameters on both the expected received signal {\em and} its statistics.
Furthermore, the analysis of the deterministic signal model in \cite{lotter20b} is extended in this work by a comprehensive steady state analysis.
This analysis reveals the impact of the number of released molecules on the steady state concentration and is crucial for the derivation of the proposed statistical model.
In summary, the model presented in this paper is a major extension of that reported in \cite{lotter20b}.

The remainder of this paper is organized as follows.
In Section~\ref{sec:system_model}, the relevant biological background is introduced and the system model is presented.
In Section~\ref{sec:deterministic_signal_model}, a deterministic model for the expected received signal in terms of an \ac{SSD} is developed.
The statistical signal model is derived in Section~\ref{sec:statistical_signal_model}.
Finally, numerical results are presented in Section~\ref{sec:results}, and our main conclusions are summarized in Section~\ref{sec:conclusion}.

\section{Biological Background and System Model}
\label{sec:system_model}
\subsection{Biological Background and Assumptions}\label{sec:system_model:assumptions}
The shapes of natural synapses are highly variable \cite{hof14}.
In this work, we adopt the cuboid domain proposed in \cite{lotter20a} as spatial model for the synaptic cleft.
Formally, it is defined in Cartesian coordinates as follows \cite{lotter20a}
\begin{align}
    \Omega = \{(x,y,z) \vert x_{\mathrm{min}} \leq x \leq x_{\mathrm{max}}, y_{\mathrm{min}} \leq y \leq y_{\mathrm{max}},\nonumber\\ z_{\mathrm{min}} \leq z \leq z_{\mathrm{max}} \}.\label{eq:domain}
\end{align}
In this model, the faces $x=\xmin$ and $x=\xmax$ represent the membranes of the presynaptic and postsynaptic neuron, respectively, and the faces in $y$ and $z$ are reflective and constrain the synaptic cleft.
By the choice of reflective faces at $y = y_{\mathrm{min}}$, $y=y_{\mathrm{max}}$, $z = z_{\mathrm{min}}$, and $z=z_{\mathrm{max}}$, the model can be applied to synapses with large extent \cite{heidelberger05} as well as to confined synaptic domains \cite{nedergaard12}.

\begin{figure*}[!t]
    \centering
    \begin{subfigure}[c]{0.49\textwidth}
        \centering
        \includegraphics[width=\textwidth]{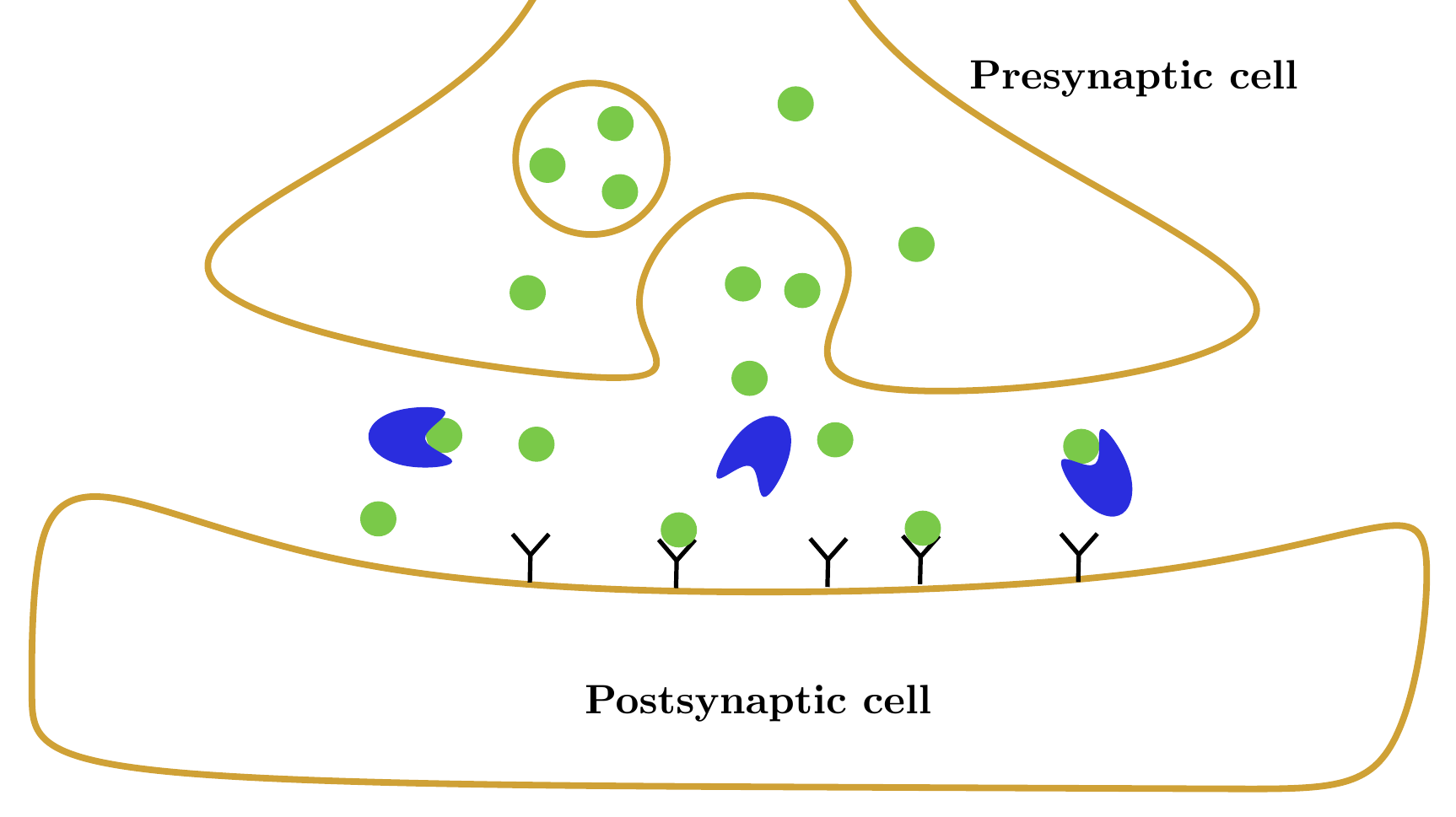}
    \end{subfigure}%
~
    \begin{subfigure}[c]{0.49\textwidth}
        \centering
        \includegraphics[width=\textwidth]{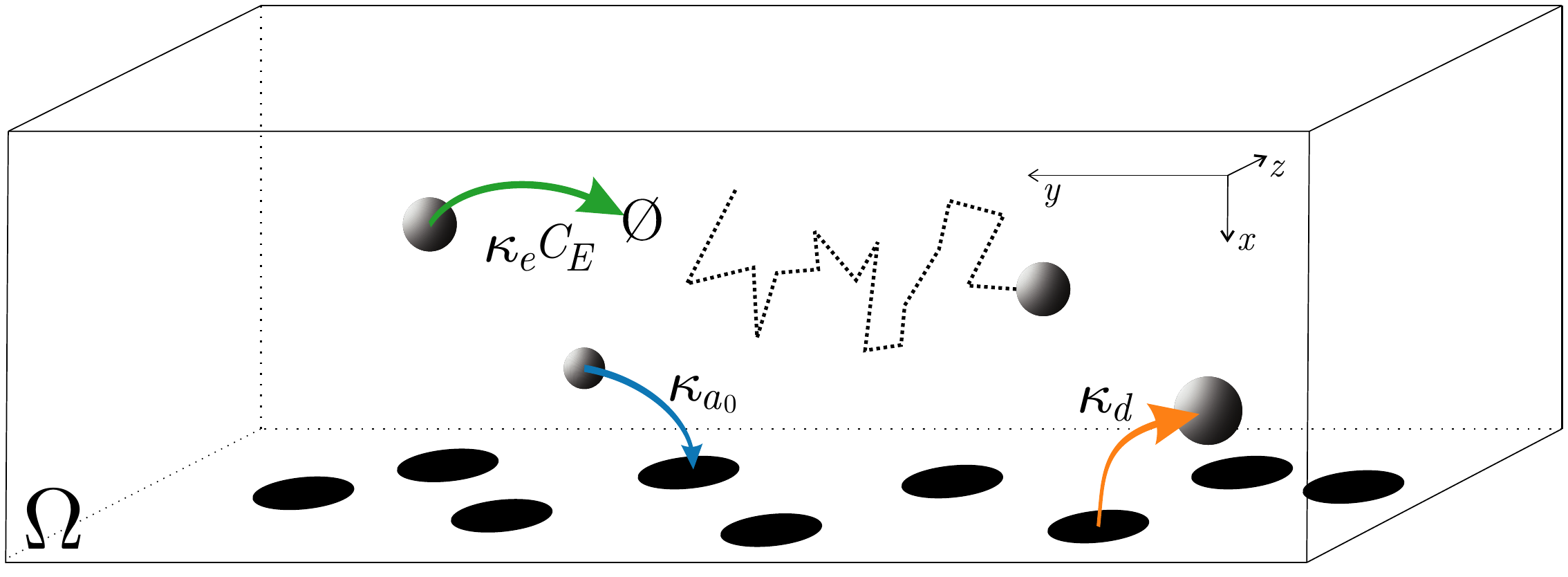}
    \end{subfigure}
    \caption{\footnotesize Model synapse. \textit{Left:} Neurotransmitters (green) enclosed in vesicles are released at the presynaptic cell, propagate by Brownian motion, and activate receptors at the postsynaptic cell. Binding to postsynaptic receptors is reversible. Solute neurotransmitters are degraded by enzymes (blue) \cite{zucker14}. \textit{Right:} Schematic diagram of the considered biological processes in domain $\Omega$ \eqref{eq:domain}, cf.~Section~\ref{sec:system_model}. Enzymatic degradation, adsorption, and desorption are shown with the associated reaction rates in green, blue, and orange, respectively. Brownian motion is depicted exemplary for one molecule (dashed line). Postsynaptic receptors are shown as black disks.}
    \label{fig:channel}
\end{figure*}

After \acp{NT} are released from presynaptic vesicles \cite{zucker14}, they propagate by Brownian motion and eventually bind to transmembrane receptors at the postsynaptic cell, the molecular receiver, where an electrical downstream signal is generated, cf.~Fig.~\ref{fig:channel}.
In nature, the residual solute \acp{NT} are either uptaken by transporter proteins at the presynaptic neuron or surrounding glial cells or degraded by enzymes to terminate synaptic signaling \cite{zucker14,bak06}.
In this way, \ac{ISI} between subsequent releases of \acp{NT} is mitigated.
While presynaptic and glial cell uptake has been considered previously by the authors (without receptor saturation) \cite{lotter20a,lotter20}, in this work, we focus on enzymatic degradation as clearance mechanism.

In the remainder of this paper, we are interested in the (random) number of molecules bound to postsynaptic receptors at time $t$, $I(t)$.

Before we state the system model, we introduce the following assumptions:
\begin{enumerate}[{A}1)]
    \item The diffusion coefficient and the total number of postsynaptic receptors are time-invariant over the time frame under consideration.\label{ass:const_rates}
    \item The diffusive propagation of molecules is fast relative to the binding to postsynaptic receptors.\label{ass:slow_reactions}
    \item Enzymatic degradation can be modeled as first-order reaction and the enzymes are uniformly distributed in space.\label{ass:enz_deg}
    \item Reversible adsorption to individual, uniformly distributed receptors with intrinsic association coefficient $\kad$ in $\si{\micro\meter\per\micro\second}$ and intrinsic dissociation rate $\kd$ in $\si{\per\micro\second}$ can be treated equivalently as reversible adsorption to a homogeneous surface with effective association coefficient $\ka$ in $\si{\micro\meter\per\micro\second}$ and dissociation rate $\kd$.\label{ass:bdr_hgz}
\end{enumerate}

A\ref{ass:const_rates} is justified if the time frame under consideration is sufficiently small.
In particular, the insertion and removal of postsynaptic receptors constitutes a long-term adaptation process \cite{gallimore16}.

A\ref{ass:slow_reactions} is plausible given experimentally observed values for the diffusion coefficient of the common \ac{NT} glutamate \cite{nielsen04} and postsynaptic receptor binding rates \cite{jonas93}.
A\ref{ass:slow_reactions} guarantees that the \acp{NT} are approximately uniformly distributed in $y$ and $z$ and, hence, the \ac{NT} concentration in $\Omega$ can be equivalently characterized by the one-dimensional \ac{NT} concentration in $\Omegax = [\xmin;\xmax]$.
We note that, because of the reflective boundaries in $y$ and $z$, $\Omega$ would in fact be equivalent to $\Omegax$ in terms of the expected number of bound receptors, if receptors did not saturate \cite{lotter20a}.
The validity of this assumption is further confirmed in Section~\ref{sec:results} by comparing the results of three-dimensional \ac{PBS} with the proposed one-dimensional model.

The first part of A\ref{ass:enz_deg} is justified in \cite{noel14} under the assumption that the degradation of molecules bound to enzymes is sufficiently fast.
The latter part is reasonable given that the enzymes exist long enough such that their concentration reaches equilibrium.
As a consequence of A\ref{ass:enz_deg}, the time constant of enzymatic degradation is given by $\kappa_e C_E$, where $C_E$ denotes the constant concentration of degrading enzymes in \si{\per\micro\meter} and $\kappa_e$ denotes the degradation rate in \si{\micro\meter\per\micro\second} \cite{noel14}.

A\ref{ass:bdr_hgz} is referred to as {\em boundary homogenization} and has been investigated in \cite{lotter20a}.
The accuracy of A\ref{ass:bdr_hgz} is further confirmed by the results presented in Section~\ref{sec:results}.

With these assumptions, we now formulate the analytical system model.

\subsection{System Model}

Based on A\ref{ass:slow_reactions} from Section~\ref{sec:system_model:assumptions}, we consider the domain $\Omegax = [\xmin;\xmax]$.

\subsubsection{Transmit Signal}
We assume a molecular transmitter, releasing \acp{NT} into the synaptic cleft instantaneously from the membrane of the presynaptic neuron $x = x_\mathrm{min}$ at time instants $\lbrace T_m \rbrace_{m \in \mathcal{M}}$, where $\emptyset \neq \mathcal{M} \subseteq \mathbb{Z}$, $\mathbb{Z}$ denotes the set of integers, $T_{m+1} > T_{m}$, and $T_m \geq 0$ denotes the (random) time of the $m$th release in \si{\micro\second}.
The transmit signal, $s(x,t)$, is then given by
\begin{align}
    s(x,t) = \sum_{m \in \mathcal{M}} N \delta(t - T_m)\delta(x),\label{eq:transmit_signal}
\end{align}
where $N$ denotes the number of released molecules per release, $\delta(\cdot)$ denotes the Dirac delta distribution, and we have assumed \ac{wlog} $\xmin = 0$.
If the presynaptic cell releases \acp{NT} at a fixed rate $1/T_s$, where $T_s$ denotes the time between two subsequent releases in \si{\micro\second}, $\mathcal{M} = \mathbb{Z}$ and $T_m$ in \eqref{eq:transmit_signal} simplifies to $m T_s$.

Instead of instantaneous release, more realistic release models which take into account the gradual release of \acp{NT} into the synaptic cleft as the vesicle holding the \acp{NT} fuses with the cell membrane have been proposed \cite{clements96}.
Here, however, for clarity of exposition, we adopt the simplified instantaneous release model which is justified if the degradation of \acp{NT} and the postsynaptic binding kinetics are relatively slow compared to the release process{\footnote{Indeed, the analysis conducted in Sections~\ref{sec:deterministic_signal_model:tfm} and \ref{sec:deterministic_signal_model:state-space_description} does not depend on the choice of $s(x,t)$, i.e., different release models can be adopted by setting $s(x,t)$ to the desired release function.}}.

\subsubsection{Channel and Expected Received Signal}\label{sec:system_model:system_model:channel:deterministic_model}

After \acp{NT} are released into the synaptic cleft, they are subject to Brownian motion, chemical reactions at the postsynaptic cell membrane, and enzymatic degradation.
Hence, on average, the particle concentration in the synaptic cleft follows the following inhomogeneous reaction-diffusion equation
\begin{align}
    \partial_t c(x,t) = D\partial_{xx} c(x,t) - \kece c(x,t) + s(x,t), \, 0 < x < a,\label{eq:diff_oned}
\end{align}
where $c(x,t)$ denotes the \ac{NT} concentration in \si{\per\micro\meter}, $D$ is the diffusion coefficient in \si{\micro\meter\squared\per\micro\second}, $\partial_t$ and $\partial_{xx}$ denote the first derivative \ac{wrt} time and the second derivative \ac{wrt} space, respectively, and we set \ac{wlog} $\xmax = a$ in \eqref{eq:domain}.

The boundary condition at the presynaptic membrane, $x=0$, is given by the no-flux boundary condition
\begin{equation}
    - D\,\partial_x c(x,t)\big\vert_{x=0} = i_x(0,t) = 0, \label{eq:pab_lb}
\end{equation}
where $\partial_x$ denotes the first derivative \ac{wrt} space and $i_x(x,t)$ denotes the particle flux in $x$ direction in \si{\per\micro\second}.

If the receptors did not saturate, the binding of \acp{NT} to postsynaptic receptors could be modeled as reversible adsorption to a homogeneous, partially absorbing boundary \cite{lotter20a}.
The corresponding boundary condition would then be \cite{lotter20a}
\begin{equation}
    i_x(a,t) = \ka c(a,t) - \kd i(t),\label{eq:pab_rb_no_sat}
\end{equation}
where $\ka$ denotes the effective adsorption coefficient in \si{\micro\meter\per\micro\second} resulting from homogenizing the postsynaptic boundary, $\kd$ denotes the dissociation rate in \si{\per\micro\second},
\begin{equation}
    i(t) = \int_{0}^{t} i_x(a,\tau) \dtau\label{eq:def:i}
\end{equation}
is the number of receptors occupied at time $t$, and we have assumed that $i(0) = 0$.

Now, the saturation of postsynaptic receptors introduces memory into the adsorption process in the sense that the rate of adsorption at time $t$ depends on $i(t)$, which in turn depends on the entire history of binding and unbinding of \acp{NT} to receptors.

Using \eqref{eq:def:i}, we propose to incorporate saturation into \eqref{eq:pab_rb_no_sat} as follows
\begin{align}
    - D\,\partial_x c(x,t)\big\vert_{x=a} &= i_x(a,t)\nonumber\\
    &= \ka \left( 1 - \frac{i(t)}{C^*} \right) c(a,t) - \kd i(t),\label{eq:pab_rb}
\end{align}
where $C^*$ denotes the total number of postsynaptic receptors.
Considering the term $\left( 1 - \frac{i(t)}{C^*} \right)$ in \eqref{eq:pab_rb}, molecules bind with the full rate $\ka$ if no receptors are occupied, i.e., $i(t)=0$, and the binding rate drops to zero, if all receptors are occupied, i.e., $i(t)=C^*$.
As $i(t)$ depends on the current and all past values of $c(a,t)$, \eqref{eq:pab_rb} is a nonlinear, state-dependent boundary condition which we refer to as {\em saturation boundary condition}.
Boundary condition \eqref{eq:pab_rb} has been proposed independently in \cite{sun20} (for an unbounded environment) and \cite{lotter20b}.

To complete the formulation of the model, we require that the initial concentration of \acp{NT} in the synaptic cleft is zero at $t=0$, i.e.,
\begin{align}
    c(x,0)=0.\label{eq:init_cond}
\end{align}
Finally, we say that a synaptic \ac{DMC} system operates in the {\em \ac{ISI}-free regime}, if there exists a small $\epsilon > 0$ such that
\begin{align}
    |T_m-T_n| > \epsilon, \, \int_{\Omegax} c(x,T_m-\epsilon) \textrm{d}x \approx 0, \, i(T_m-\epsilon) \approx 0,\label{eq:isi_free_regime}
\end{align}
for all $m,n \in \mathcal{M}$ in \eqref{eq:transmit_signal} and $m \neq n$.
Eq.~\eqref{eq:isi_free_regime} implies that the synaptic cleft is cleared of \acp{NT} between any two releases of \acp{NT}\footnote{Since the total number of solute and bound \acp{NT} does not increase between two subsequent releases of \acp{NT}, it is sufficient to require \eqref{eq:isi_free_regime} for one specific $\epsilon$.}.

\subsubsection{Received Signal}\label{sec:system_model:system_model:statistics}
The deterministic model introduced in Section \ref{sec:system_model:system_model:channel:deterministic_model} yields the {\em expected} number of bound (activated) postsynaptic receptors.
The {\em actual} number of bound receptors, however, is subject to random fluctuations.
This randomness originates from four sources.
\begin{enumerate}[{R}1)]
    \item The {\em propagation} of particles due to Brownian motion is random.
    \item There is randomness in the {\em enzymatic degradation} of particles, i.e., the time at which a particle is degraded is random.
    \item When an \ac{NT} hits a receptor at the postsynaptic side, the {\em actual binding} is random and dependent on the receptor state.
    When the receptor is unbound, the binding probability is determined by the intrinsic absorption coefficient of the receptor.
    When the receptor is bound, the binding probability is $0$.
    Now, the receptor state itself is the result of a series of random events (the binding and unbinding events of all molecules).
    Hence, the random binding of one molecule to a postsynaptic receptor depends on all previous random binding events of all other molecules.
    \item The {\em dissociation} (unbinding) of \acp{NT} from receptors is random.
\end{enumerate}

If there is only one single release of $N$ molecules and if the fate of each molecule is independent of the other molecules, the number of bound molecules at time $t$, $I(t)$ can be modeled as binomial distributed random variable, $I(t) \sim B(n;i(t)/N,N)$, where $B(n;p',N')$ denotes the binomial distribution with parameters $p'$ and $N'$ \cite{jamali19}.
Similarly, if the concentration of solute molecules is very large compared to the total number of receptors $C^*$, assuming statistical independence of the receptors, $I(t)$ can be approximated as $I(t) \sim B(n;i(t)/C^*,C^*)$ \cite{kuscu18}.
However, the assumption of statistical independence of {\em molecules} is violated when {\em molecules compete for receptors}, i.e., when the probability that a molecule binds to a receptor depends on how many receptors are already occupied by other molecules.
On the other hand, statistical independence of {\em receptors} cannot be assumed when {\em receptors compete for molecules}.
This is the case if relatively few molecules are available such that the trapping of molecules at receptors significantly alters the molecule concentration and thereby impacts the binding probability of the unbound receptors.
Hence, in both of these cases the considered events are statistically dependent.

According to these observations and since the concentration of \acp{NT} in the synaptic cleft is highly variable \cite{zucker14}, statistical independence can in general be neither assumed between molecules nor between receptors.
However, the competition of receptors for \acp{NT} is primarily relevant if relatively few molecules are exposed to the receptors, i.e., if few molecules are released and the contribution of residual \acp{NT} due to \ac{ISI} is negligible, cf.~\eqref{eq:isi_free_regime}.
If many \acp{NT} are present due to a large number of released molecules or due to \ac{ISI}, on the other hand, the assumption of statistical independence between receptors is satisfied and the binomial model $I(t) \sim B(n;i(t)/C^*,C^*)$ applies.

Let us denote the probability that $n$ receptors are bound at time $t$ as $P_I(n)$, i.e., $I(t) \sim P_I(n)$\footnote{The dependence of $P_{I}$ on $t$ is only implicit in the notation but clear from the definition.}.
In Section~\ref{sec:statistical_signal_model:single_release}, we derive an analytical expression for $P_I(n)$ in terms of the hypergeometric distribution which captures the statistical dependence between receptors and \acp{NT}, respectively, in the \ac{ISI}-free case.
By encompassing this regime, the proposed statistical model complements the existing binomial models and constitutes a step towards a comprehensive signal model in the presence of competition between receptors or \acp{NT}, respectively.

\section{Deterministic Signal Model}
\label{sec:deterministic_signal_model}
In this section, we first derive the equilibrium number of bound \acp{NT} after a finite number of instantaneous releases in the absence of enzymatic degradation.
Then, we reformulate \labelcref{eq:diff_oned,eq:pab_lb,eq:pab_rb,eq:init_cond} in terms of a \ac{TFM} and finally derive a deterministic signal representation in terms of an \ac{SSD}.

\subsection{Steady State in the Absence of Enzymatic Degradation}\label{sec:deterministic_signal_model:steady_state}
We consider $\mathcal{M} \subset \mathbb{Z}$ in \eqref{eq:transmit_signal} finite, i.e., $|\mathcal{M}|<\infty$, and are interested in $i^{\infty} = \lim\limits_{t \to \infty} i(t)$.

\begin{theorem}\label{thm:steady_state}
Let $\kece=0$, $\ka,\kd > 0$, and $|\mathcal{M}|<\infty$. Then $i^\infty = \lim\limits_{t \to \infty} i(t)$ is the smallest positive solution to the quadratic equation
\begin{align}
    \left(i^{\infty}\right)^2 -\left[\left(1+\frac{a \kd}{\ka}\right)C^* + N |\mathcal{M}|\right] i^{\infty} + N |\mathcal{M}| C^* = 0,\label{eq:steady_state}
\end{align}
i.e.,
\begin{align}
    i^\infty = \frac{1}{2} &\left\lbrace\left(1+\frac{a \kd}{\ka}\right)C^* + N |\mathcal{M}|\right.\nonumber\\
    &-\left.\sqrt{\left[\left(1+\frac{a \kd}{\ka}\right)C^* + N |\mathcal{M}| \right]^2 - 4 N |\mathcal{M}| C^*}\right\rbrace.\label{eq:i_infty}
\end{align}
\end{theorem}
\begin{IEEEproof}
    Please see Appendix~\ref{sec:app:proof_thm_1}.
\end{IEEEproof}
\begin{remark}
As $C^* \to \infty$, \eqref{eq:steady_state} is dominated by the terms $-\left(1+\frac{a \kd}{\ka}\right)C^* i^{\infty} + N |\mathcal{M}| C^* = 0$.
In this case, dividing by $C^*$, we obtain 
\begin{align}
    \lim\limits_{C^* \to \infty} i^{\infty} = N |\mathcal{M}| \frac{\ka}{\ka + a \kd},
\end{align}
which corresponds to the steady state without receptor saturation obtained in \cite{lotter20a}.
Hence, for infinitely many receptors, the non-saturating linear model is recovered.
\end{remark}
We conclude this subsection observing that, after dividing \eqref{eq:steady_state} by $N |\mathcal{M}|$,
\begin{align}
    \lim\limits_{N|\mathcal{M}| \to \infty} i^{\infty} = C^*.
\end{align}
Hence, if the number of emitted molecules is large enough, all receptors are eventually bound.

\subsection{Transfer Function Model}\label{sec:deterministic_signal_model:tfm}
In this section, we derive a deterministic signal model in terms of transfer functions by transformation of the boundary-value problem from Section~\ref{sec:system_model:system_model:channel:deterministic_model}.
We first consider the special case $\ka=\kd=\kece=0$ for which the system model \eqref{eq:diff_oned},\eqref{eq:pab_lb},\eqref{eq:pab_rb},\eqref{eq:init_cond} reduces to a linear system and the corresponding \ac{SSD} can be readily obtained.
This model is extended in Section~\ref{sec:deterministic_signal_model:state-space_description} to account for saturation, desorption, and enzymatic degradation, i.e., $\ka,\kd,\kece \geq 0$.

\subsubsection{Vector Representation}
Assuming $\kece=0$, the partial differential equation \eqref{eq:diff_oned} is decomposed into a continuity equation and a concentration gradient as follows
\begin{align}
	\partial_t c(x,t) &= -\partial_x i_x(x,t) + s(x,t), && 0 < x < a,
	\label{eq:diff:cont}\\
	i_x(x,t) &= -D\partial_{x} c(x,t), && 0 < x < a.
	\label{eq:diff:grad}
\end{align}
For the derivation of a \ac{TFM}, Eqs.~\eqref{eq:diff:cont}, \eqref{eq:diff:grad} are arranged into vector form \cite[Eq.~(12)]{schaefer:laminar:2020}
\begin{align}
    &\left[\bm{D}\partial_t - \bm{\mathrm{L}}\right]\bm{y}(x,t) = \bm{f}(x,t), \label{eq:tf:1}
\end{align}
with capacitance matrix $\bm{D}\in\mathbb{R}^{2\times 2}$ and the $2\times 2$ matrix-valued spatial differential operator $\bm{\mathrm{L}}$
\begin{align}
    \bm{D} &= \begin{bmatrix}
        0 & 0\\
        1 & 0
    \end{bmatrix}, 
    &\bm{\mathrm{L}} = -\begin{bmatrix}
        \partial_x & \nicefrac{1}{D} \\
        0 & \partial_x
    \end{bmatrix}.
    \label{eq:tf:2}
\end{align}
The physical quantities are arranged in the vector $\bm{y}\in\mathbb{R}^{2\times 1}$ and vector $\bm{f}\in\mathbb{R}^{2\times 1}$ contains the transmit signal $s$ from \eqref{eq:transmit_signal}
\begin{align}
    \bm{y}(x,t) &= 
    \begin{bmatrix}
        c(x,t)&
        i_x(x,t)
    \end{bmatrix}\tran\!\!, 
    &\bm{f}(x,t) = 
    \begin{bmatrix}
        0&
        s(x,t)
    \end{bmatrix}\tran,
    \label{eq:tf:3}
\end{align}
where $(\cdot)\tran$ denotes transposition.
Eqs.~\eqref{eq:pab_lb}, \eqref{eq:pab_rb} are represented with the boundary operator $\Fbh\in\mathbb{R}^{2\times2}$, acting on $\bm{y}(x,t)$.
This yields the vector $\bm{\phi}\in\mathbb{R}^{2\times 1}$ of boundary values as follows
\begin{align}
    &\Fbh\bm{y}(x,t) = \bm{\phi}(x,t), &x = 0,a,
    \label{eq:tf:4}
\end{align}
where $\Fbh$ and the vectorized boundary values $\bm{\phi}(0,t)$ and $\bm{\phi}(a,t)$ are defined as follows
\begin{align}
    &\Fbh = \begin{bmatrix}
        0 & 0\\
        0 & 1
    \end{bmatrix},
    & \bm{\phi}(0,t) = \bm{0},
    &&\bm{\phi}(a,t) = \begin{bmatrix}
        0\\
        \phi_i(t)
    \end{bmatrix}.
    \label{eq:tf:5}
\end{align}
Boundary value $\phi_i(t)$ in \eqref{eq:tf:5} is used as placeholder for the right-hand side of the nonlinear boundary condition \eqref{eq:pab_rb}.  
In the presented form, Eqs.~\eqref{eq:tf:1}, \eqref{eq:tf:4} represent a one-dimensional diffusion process with Neumann boundary conditions. For the special case $\ka=\kd=\kece=0$ considered in this section, the boundary condition \eqref{eq:pab_rb} reduces to a homogeneous boundary, i.e., the boundary value $\phi_i(t)$ is equal to zero for all $t$.
However, $\phi_i(t)$ is kept as a placeholder for the general case $\ka,\,\kd,\,\kece\geq0$ that is investigated in Section~\ref{sec:deterministic_signal_model:state-space_description}.

\subsubsection{Functional Transformations}

To reduce the number of independent variables in \eqref{eq:tf:1}, \eqref{eq:tf:4}, $\bm{y}(x,t)$ is expanded in terms of an infinite set of bi-orthogonal eigenfunctions $\Kprim_\mu(x) \in\mathbb{R}^{2\times 1}$ and $\Kadj_\mu(x) \in\mathbb{R}^{2\times 1}$ of spatial differentiation operator $\bm{\mathrm{L}}$.
The corresponding eigenvalues $s_\mu$ define the discrete spectrum of $\bm{\mathrm{L}}$ \cite{churchill:1972}.
Both, eigenvalues and eigenfunctions are indexed with $\mu\in\mathbb{N}_0$, where $\mathbb{N}_0$ denotes the set of non-negative integers.
With the eigenfunctions $\Kprim_\mu$, $\Kadj_\mu$ not yet determined, a forward and inverse Sturm-Liouville transformation (SLT) is defined as follows \cite{churchill:1972}
\begin{align}
	\bar{y}_\mu(t) &= \int_{0}^{a}\Kadj_\mu\tran(x)\bm{D}\bm{y}(x,t)\dint{x},\label{eq:for_slt}\\
	\bm{y}(x,t) &= \sum_{\mu = 0}^{\infty} \frac{1}{N_\mu}\bar{y}_\mu(t) \Kprim_\mu(x). 
	\label{eq:for_inv_slt}
\end{align}
The forward SLT \eqref{eq:for_slt} expands the vector $\bm{y}$ into the eigenfunctions $\Kadj_\mu$ yielding the expansion coefficients $\bar{y}_\mu$. The inverse SLT \eqref{eq:for_inv_slt} represents $\bm{y}$ by a series expansion with eigenfunctions $\Kprim_\mu$ and scaling factors $N_\mu$. 
The exact form of the eigenfunctions $\Kprim_\mu$ can be derived from the following eigenvalue problem with homogeneous boundary conditions \cite{schaefer:nds:2017a}
\begin{align}
	\bm{\mathrm{L}}\Kprim_\mu(x) &= s_\mu\bm{D}\Kprim_\mu(x), &&0<x<a, \label{eq:evK}\\
	\Fbh\Kprim_\mu(x) &= \bm{0}, && x= 0,a. \label{eq:evFb}
\end{align}
The spatially one-dimensional eigenvalue problem \eqref{eq:evK} can be solved for the eigenfunctions $\Kprim_\mu$ in terms of a matrix exponential \cite{schaefer:nds:2017a}
\begin{align}
	&\Kprim_\mu(x) = \mathrm{e}^{\bm{Q}_\mu\,x}\Kprim_\mu(0),
	&\bm{Q}_\mu = -\begin{bmatrix}
		0 & \nicefrac{1}{D}\\
		s_\mu & 0
	\end{bmatrix}, \label{eq:ev_mat}
\end{align}
where matrix $\bm{Q}_\mu$ can be derived from \eqref{eq:evK} with $\bm{D}$ and $\bm{\mathrm{L}}$ from \eqref{eq:tf:2}. 

The eigenfunctions $\Kadj_\mu$ can be derived from an eigenvalue problem similar to \eqref{eq:evK} (see \cite{rabenstein18}), or from the following relation, $\Kadj_\mu(x) = \left(\mathrm{e}^{-\bm{Q}_\mu\,x}\right)\tran\Kadj_\mu(0)$ \cite[Eq.~(104)]{rabenstein18}. 
From the calculation of the matrix exponential in \eqref{eq:ev_mat} with any suitable method, e.g., the procedure in \cite[Sec.~III]{schaefer:nds:2017a}, and after the evaluation of \eqref{eq:evFb} at $x = 0$, the eigenfunctions $\Kprim_\mu(x)$ and $\Kadj_\mu(x)$ follow as
\begin{align}
    &\Kprim_\mu(x) \!=\! \begin{bmatrix}
        \cos(\gamma_\mu x)\\
        D\gamma_\mu\sin(\gamma_\mu x)
    \end{bmatrix}\!\!,\!\!\!\!
    &\Kadj_\mu(x) \!=\! \begin{bmatrix}
        -D \gamma_\mu \sin(\gamma_\mu x)\\
        \cos(\gamma_\mu x) 
    \end{bmatrix}\!\!.
    \label{eq:tf:6} 
\end{align}
The eigenvalues $s_\mu$ and wavenumbers $\gamma_\mu$ can be derived from \eqref{eq:tf:6} together with the boundary conditions \eqref{eq:evFb} at $x = a$ as $s_\mu = -D\,\gamma_\mu^2$, $\gamma_\mu = \mu \frac{\pi}{a}$ \cite[Sec.~IV]{schaefer:nds:2017a}.
Finally, to ensure the existence of the inverse transformation in \eqref{eq:for_inv_slt}, the eigenfunctions in \eqref{eq:tf:6} have to be bi-orthogonal, yielding the following expression for the scaling factor \cite{churchill:1972}
\begin{align}
    N_\mu = \int_0^a\Kadj_\mu\tran(x)\bm{D}\Kprim_\mu(x)\dint{x} = \begin{cases}
        a & \mu = 0\\
        \nicefrac{a}{2} & \mu \neq 0
    \end{cases}.
    \label{eq:tf:8}
\end{align}
The application of the forward transformation \eqref{eq:for_slt} to \eqref{eq:tf:1} leads to the expansion coefficients
\begin{align}
    \bar{y}_\mu(t) &= \mathrm{e}^{s_\mu t} \overset{t}{*}\left(\bar{f}_\mu(t) - \bar{\phi}_\mu(t)\right),\label{eq:tf:10}
\end{align}
where $\overset{t}{*}$ denotes convolution \ac{wrt} time, and $\bar{f}_\mu(t)$ and $\bar{\phi}_\mu(t)$ follow from the expansion of $\bm{f}(x,t)$ in \eqref{eq:tf:3} and $\bm{\phi}(x,t)$ in \eqref{eq:tf:5} as
\begin{align}
    &\bar{f}_\mu(t) \!=\!\!\int_0^a\!\!\!\Kadj_\mu\tran(x)\bm{f}(x,t)\dint{x},\!\!\!
    &\bar{\phi}_\mu(t) \!=\! \left[\Kadj_\mu\tran(x) \bm{\phi}(x,t)\right]_0^a\!\!\!\quad, \label{eq:tf:11}
\end{align}
where $[f(x)]_a^b = f(b) - f(a)$.
For the numerical evaluation, the infinite sum in \eqref{eq:for_inv_slt} is truncated to $\mu = 0, \dots, Q-1$.
The accuracy of the computed solution hence depends on $Q$.

\subsubsection{State-Space Description}
We conclude this section by deriving an equivalent formulation of \eqref{eq:tf:10}, \eqref{eq:for_inv_slt} in terms of an \ac{SSD}.
The state-space representation will subsequently allow us in Section~\ref{sec:deterministic_signal_model:state-space_description} to incorporate the nonlinear effects due to receptor saturation that we have avoided in this section by setting $\ka=0$.
To derive the \ac{SSD}, the expansion coefficients $\bar{y}_\mu$ in \eqref{eq:tf:10} and the synthesis equation \eqref{eq:for_inv_slt} for $\bm{y}$ are transformed into discrete time using an impulse invariant transform \cite{schaefer:laminar:2020}. This yields the following discrete-time \ac{SSD} \cite{schaefer:ecc:2019}
\begin{align}
    \bar{\bm{y}}[k+1] &= \mathrm{e}^{\As T}\bar{\bm{y}}[k] +T\bar{\bm{f}}[k+1] - T\bar{\bm{\phi}}[k+1], \label{eq:tf:12}\\
    \bm{y}[x,k] &= \Cs(x)\bar{\bm{y}}[k], \label{eq:tf:13}
\end{align}
with discrete-time index $k$ and sampling interval $T$, i.e., $t = kT$.
Consequently, $\bm{y}[x,k]$, $c[x,k]$, and $i_x[x,k]$ are given by $\bm{y}(x,kT)$, $c(x,kT)$, and $i_x(x,kT)$, respectively.
The sampling interval $T$ should be adapted to the smoothness of $\bm{y}(x,t)$ to ensure that $\bm{y}(x,t)$ is accurately reproduced by $\bm{y}[x,k]$. Specifically, the smoother signal $\bm{y}(x,t)$ is, the larger may $T$ be chosen. $\bm{y}(x,t)$ in turn is the smoother, the smaller $D$, $\kad$, $\kd$, and $\kece$ are.
For a numerical example of how to choose $T$, please see Table~\ref{tab:sim_params}.
In the discrete-time SSD in \eqref{eq:tf:12} and \eqref{eq:tf:13}, \textit{state equation} \eqref{eq:tf:12} is the vector-valued discrete-time equivalent of \eqref{eq:tf:10}, where vector $\bar{\bm{y}}\in\mathbb{R}^{Q\times 1}$ contains $Q$ coefficients $\bar{y}_\mu$ and diagonal matrix $\As\in\mathbb{R}^{Q\times Q}$ contains $Q$ eigenvalues $s_\mu$ on its main diagonal, i.e., $\bar{\bm{y}}[k]=[\bar{y}_{0}(kT),\ldots,\bar{y}_{Q-1}(kT)]\tran=\left(\bar{y}_{\mu}(kT)\right)_{\mu=0}^{Q-1}$ and $\As \!=\! \mathrm{diag}\left\lbrace s_0, \dots, s_{Q-1}\right\rbrace$.
Vectors $\bar{\bm{f}}[k]\in\mathbb{R}^{Q\times 1}$ and $\bar{\bm{\phi}}[k]\in\mathbb{R}^{Q\times 1}$ are defined as $\bar{\bm{f}}[k] = \left(\bar{f}_{\mu}(kT)\right)_{\mu=0}^{Q-1}$ and $\bar{\bm{\phi}}[k] = \left(\bar{\phi}_{\mu}(kT)\right)_{\mu=0}^{Q-1}$, respectively, and contain $Q$ values of $\bar{f}_{\mu}$ and $\bar{\phi}_{\mu}$ from \eqref{eq:tf:11}.
\textit{Output equation} \eqref{eq:tf:13} is the discrete-time equivalent of the synthesis equation in \eqref{eq:for_inv_slt}, where the summation is replaced by a multiplication with matrix $\Cs(x)\in\mathbb{R}^{2\times Q}$,
\begin{align}
    \Cs(x) &= \left[\nicefrac{1}{N_0}\Kprim_0(x), \dots, \nicefrac{1}{N_{Q-1}}\Kprim_{Q-1}(x) \right].\label{eq:tf:15}
\end{align}
For the following steps, we further define matrix $\tilde{\Cs}(x)\in\mathbb{R}^{2\times Q}$,
\begin{align}
    \Csa(x) &= \left[\Kadj_0(x), \dots, \Kadj_{Q-1}(x) \right].	\label{eq:tf:16}
\end{align}

\subsection{Receptor Saturation, Enzymatic Degradation, and Number of Molecules}\label{sec:deterministic_signal_model:state-space_description}
In this section, we consider the general case $\ka,\kd,\kece \geq 0$.
To this end, we use feedback loops to incorporate the saturation boundary condition \eqref{eq:pab_rb} into the \ac{SSD} derived in the previous section.
Finally, we also incorporate enzymatic degradation in state-space and use the resulting \ac{SSD} to derive the total number of molecules in the system at time $t$.

\subsubsection{Incorporation of Receptor Saturation}
\label{subsec:sat}
First, we incorporate the effect of receptor saturation and desorption into the discrete-time model \eqref{eq:tf:12}, \eqref{eq:tf:13} by applying boundary condition \eqref{eq:pab_rb}. 
To this end, we choose the placeholder boundary value $\phi_i(t)$ introduced in \eqref{eq:tf:5} as the modified discrete-time equivalent of boundary condition \eqref{eq:pab_rb}
\begin{align}
    \phi_i[k+1] = i_x[a,k+1] = \hat{\kappa}_a[k]\, c[a,k] - \hat{\kappa}_d[k], \label{eq:tf:17}
\end{align}
where the discrete-time reaction rates are defined as
\begin{align}
    &\hat{\kappa}_a[k] = \kappa_a\left(1 - \frac{i[k]}{C^*}\right), 
    &\hat{\kappa}_d[k] = \kappa_d\,i[k],  \label{eq:tf:18}
\end{align}
and the discrete-time equivalent $i[k]$ of the accumulated net flux in \eqref{eq:def:i} is defined as
\begin{align}
    i[k] = T\sum_{k' = 0}^{k}i_x[a,k']. \label{eq:tf:19}
\end{align}
In \eqref{eq:tf:17}, the value of $\phi_i$ in time slot $k+1$ depends on the values of the concentration $c$ and flux $i$ in time slot $k$.
Hence, \eqref{eq:tf:17} introduces a delay of $T$ in the computation of the flux compared to the right-hand side of \eqref{eq:pab_rb}.
The technical rationale behind this delay is to avoid a delay-free loop in the \ac{SSD}, cf.~Fig.~\ref{fig:block_ssd}, which would render the system unstable.
However, if $T$ is chosen small enough relative to the velocity of the binding kinetics defined by $\kappa_a$ and $\kappa_d$, the delay is justified physically, because in this case, $i$ and $c$ are approximately constant in two subsequent sampling intervals.
We verify the accuracy of this assumption in Section~\ref{sec:results} with \ac{PBS}.

In order to incorporate \eqref{eq:tf:17} into the discrete-time \ac{SSD}, vector $\bar{\bm{\phi}}$ in \eqref{eq:tf:12} is computed using the vector-valued discrete-time version of \eqref{eq:tf:11}
\begin{align}
    \bar{\bm{\phi}}[k+1] &= \left[\Csa\tran(x) \bm{\phi}(x,(k+1)T)\right]_0^a. \label{eq:tf:20}
\end{align}
Exploiting the structure of $\bm{\phi}(x,t)$ in \eqref{eq:tf:5}, $\tilde{\Cs}(x)$ in \eqref{eq:tf:16}, and the definition of $\phi_i$ in \eqref{eq:tf:17}, we can write \eqref{eq:tf:20} as follows
\begin{align}	
    \bar{\bm{\phi}}[k+1] &=\tilde{\bm{c}}_2(a)p[k+1]
    = \tilde{\bm{c}}_2(a)\hat{\kappa}_a[k]\, c[a,k] - \tilde{\bm{c}}_2(a)\hat{\kappa}_d[k], \label{eq:tf:21}
\end{align}
where $\tilde{\bm{c}}_2(x)\in\mathbb{R}^{Q\times 1}$ is the second column of $\Csa\tran(x)$ in \eqref{eq:tf:16} containing the second entries $\tilde{K}_{2,\mu}(x) = \cos(\gamma_\mu x)$ of $\Kadj_\mu$ in \eqref{eq:tf:6}, i.e., $\tilde{\bm{c}}_2(x) = \left(\tilde{K}_{2,\mu}(x)\right)_{\mu=0}^{Q-1}$.
Furthermore, concentration $c$ can be expressed as follows according to output equation \eqref{eq:tf:13}
\begin{align}
    c[a,k] = \bm{c}_1\tran(a)\bar{\bm{y}}[k], \label{eq:tf:22}
\end{align}
where $\bm{c}_1\tran(x)\in\mathbb{R}^{1\times Q}$ is the first row of matrix $\Cs(x)$ in \eqref{eq:tf:15} containing the first entries $K_{1,\mu}(x) = \cos(\gamma_\mu x)$ of $\Kprim_\mu$ in \eqref{eq:tf:6}, i.e., $\bm{c}_1(x) = \left(K_{1,\mu}(x)\right)_{\mu=0}^{Q-1}$.
Inserting \eqref{eq:tf:22} into \eqref{eq:tf:21} leads to
\begin{align}
    \bar{\bm{\phi}}[k+1] &= \hat{\kappa}_a[k]\Ks_a\bar{\bm{y}}[k] -\hat{\kappa}_d[k] \Ks_d, \label{eq:tf:23}
\end{align}
with matrix $\Ks_a = \tilde{\bm{c}}_2(a)\bm{c}_1\tran(a)$ and vector $\Ks_d = \tilde{\bm{c}}_2(a)$.
Inserting \eqref{eq:tf:23} into \eqref{eq:tf:12}, we obtain the state equation for $\ka,\kd \geq 0$, that accounts for receptor saturation and desorption at $x = a$
\begin{align}
    \bar{\bm{y}}[k+1] &= \left(\mathrm{e}^{\As T} - T\hat{\kappa}_a[k]\Ks_a\right)\bar{\bm{y}}[k] \nonumber  \\  &\quad\quad\quad + T\hat{\kappa}_d[k]\Ks_d+ T\bar{\bm{f}}[k+1]. \label{eq:tf:25}
\end{align}

\subsubsection{Incorporation of Degradation}
To complete the model, the enzymatic degradation in \eqref{eq:diff_oned} has to be incorporated into \eqref{eq:tf:25}.
The degradation reaction is modeled as a first-order reaction, cf.~A\ref{ass:enz_deg} in Section~\ref{sec:system_model}, and it can be incorporated into \eqref{eq:tf:25} using a decaying exponential function $\mathrm{e}^{-\kece t}$ \cite{noel14}.
This yields the following discrete-time model
\begin{align}
    \bar{\bm{y}}[k+1] &= \left(\mathrm{e}^{-\kece T}\mathrm{e}^{\As T} - T\hat{\kappa}_a[k]\Ks_a\right)\bar{\bm{y}}[k] \nonumber \\  &\quad\quad\quad+ T\hat{\kappa}_d[k] \Ks_d + T\bar{\bm{f}}[k+1]. \label{eq:tf:26} 
\end{align}
The modified state equation \eqref{eq:tf:26} accounts for saturation and desorption at $x = a$ according to \eqref{eq:pab_rb} and enzymatic degradation, while the output equation \eqref{eq:tf:13} to calculate the NT concentration and flux remains unchanged.
We note that \eqref{eq:tf:26} collapses to \eqref{eq:tf:12} if $\ka=\kd=\kece=0$.
Figure~\ref{fig:block_ssd} shows the block diagram of state equation \eqref{eq:tf:26} including the effects of saturation (blue), desorption (orange) and degradation (green), which are incorporated using a feedback structure.

\begin{figure}[!tbp]
	\centering
    \scalebox{.85}{
	\begin{tikzpicture}[node distance = 1.5cm, rounded corners = 2pt]
		\node[dspsquare, rounded corners = 2pt](del) at (0,0){$z^{-1}$};
		\node[dspsquare, minimum width=1.3cm](A) at (1,-1) {$\mathrm{e}^{\As T}$};
		\node[dspsquare, fill = mygreen!80, minimum width=1.4cm](deg) at (-1.1,-1) {$\mathrm{e}^{-\kappa_eC_e T}$};
		\node[dspsquare, fill = myblue!80,minimum width=1.5cm](absT) at (1.5,-2.5) {$\hat{\kappa}_\mathrm{a}[k]$};
		\node[dspsquare, fill = myblue!80, minimum width=1.3cm](absMat) at (-1,-2.5) {$T \Ks_a$};
		\node[dspsquare, fill = myorange!80, minimum width=1.5cm](desT) at (1.5,-4.5) {$\hat{\kappa}_\mathrm{d}[k]$};
		\node[dspsquare, fill = myorange!80, minimum width=1.3cm](desMat) at (-1,-4.5) {$T \Ks_d$};
		\node[dspsquare, right of = del, node distance = 3.5cm, minimum width=1.3cm](ca) {$\bm{c}_1\tran(a)$};
		
		\node[dspadder, left of = del, node distance = 2.5cm](add1){};
		\node[dspadder, below of = add1, node distance = 1cm](add2){};
		\node[dspadder, below of = add2, node distance = 1.5cm](add3){};
		
		\node[dspnodeopen, left of = add1, node distance = 1cm](fe) {$T \bar{\bm{f}}[k+1]$}; 
		\node[dspnodefull, right of = del, node distance = 2cm](n1){};
		\node[coordinate](c1) at (1.75,-1.75) {};
		\node[dspmultiplier, right of = absT, node distance = 3cm](mul1) {};
		\node[dspmultiplier, left of = absT, node distance = 1.25cm](mul2) {};
		\node[dspadder, below of = mul1, node distance = 1cm](add4) {};
		\node[dspnodefull, left of = add4, node distance = 2cm](n3) {};
        \node[inner sep = 0pt, left of = n3, node distance=1.75cm](label){Eq.~\eqref{eq:tf:18}};
		\node[dspsquare, rounded corners = 2pt, right of = absT, node distance = 2cm](del2){$z^{-1}$};
		\node[dspsquare, rounded corners = 2pt, below of = del2, node distance = 1.75cm](del3){$z^{-1}$};
		\node[dspsquare, rounded corners = 2pt, below of = del2, node distance = 2.75cm](del4){$z^{-1}$};
   		\node[dspsquare, rounded corners = 2pt, above of = mul1, node distance = 1.5cm](del5){$z^{-1}$};
		\node[coordinate, right of = del3, node distance = 1.5cm](c2){};
		\node[inner sep = 0pt, above of = add4, node distance = 1cm](x){$\times$};
		\node[inner sep = 0pt, left of = absT, node distance = 1.25cm](x2){$\times$};
		\node[inner sep = 0pt, above right = -0.01cm and 0.01cm of add4](x){$-$};
		\node[inner sep = 0pt, above right = -0.01cm and 0.01cm of add3](x){$-$};
		
		\draw[dspconn] (fe) -- (add1); 
		\draw[dspflow] (add1) -- node[above, midway]{$\bar{\bm{y}}[k+1]$} (del);
		\draw[dspflow] (del) -- node[above, midway]{$\bar{\bm{y}}[k]$} (n1);
		\draw[dspflow] (n1) -- (ca);
   		\draw[dspconn] (ca) -| node[midway,right]{$c[a,k]$} (del5);
  		\draw[dspconn] (del5) -- (mul1);
		\draw[dspconn] (mul1) -- (add4); 
		\draw[dspflow] (add4) -- node[midway, above]{$i[k]$}(n3); 
		\draw[dspline] (n3) |- (del3);
		\draw[dspconn] (del3) -| (add4);
		\draw[dspconn] (n3) -| (absT);
		\draw[dspconn] (n3) -| (desT);
		\draw[dspflow] (n1) |- (A);
		\draw[dspconn] (A) -- (deg) -- (add2);
		\draw[dspconn] (add2) -- (add1);
		\draw[dspconn] (add3) -- node[midway, right]{$\bar{\bm{\phi}}[k+1]$}(add2);
		\draw[dspline] (n1) |- (c1);
		\draw[dspconn] (c1) -| (mul2);
		\draw[dspconn] (absT) -- (mul2);
		\draw[dspconn] (mul2)  -- (absMat) -- (add3);
		\draw[dspconn] (absT) -- (del2) -- (mul1);
		\draw[dspconn] (desT) -- (desMat);
		\draw[dspconn] (desMat)  -| (add3);
		\draw[dspflow] (desT) |- (del4);
		\draw[dspline] (del4) -| (c2);
		\draw[dspconn] (c2) |- (add4);				
	\end{tikzpicture}
    }
	\caption{Block diagram of the proposed discrete-time SSD in \eqref{eq:tf:26}, including saturation (blue), desorption (orange), and degradation (green). The time-variant coefficients $\hat{\kappa}_a$ and $\hat{\kappa}_d$ are computed using \eqref{eq:tf:18} and the accumulated net flux $i$ in \eqref{eq:tf:19}. }
	\label{fig:block_ssd}
\end{figure}
\begin{figure}[!t]
    \centering
    \includegraphics[width=\linewidth]{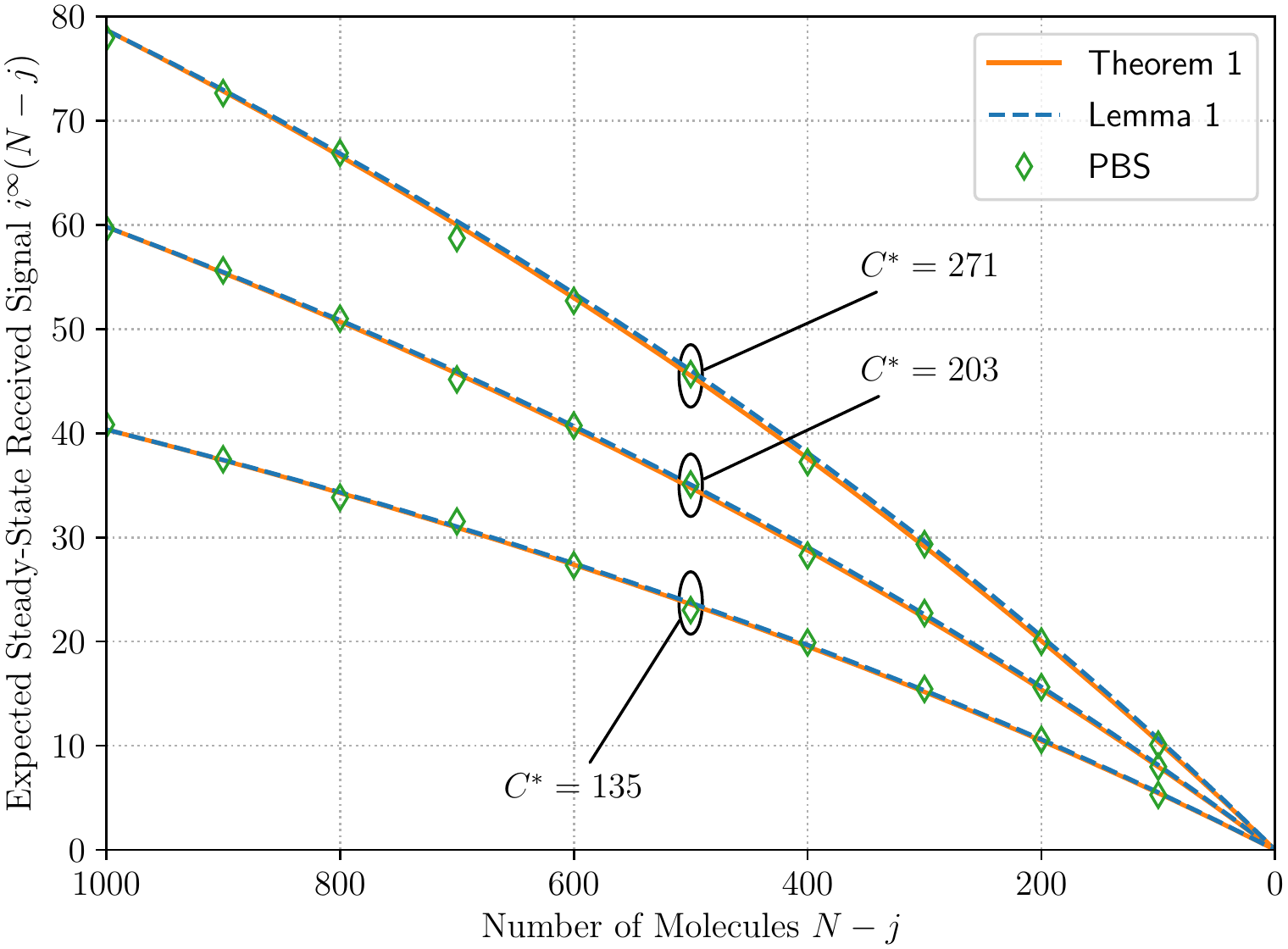}
    \caption{Impact of the number of released molecules on the expected number of bound molecules in the steady state for different numbers of postsynaptic receptors. The analytically obtained exact value from Theorem~\ref{thm:steady_state} (orange line) is compared to the approximation proposed in Lemma~\ref{lem:steady_state_approximation} (blue line) and results from \ac{PBS} (green markers), cf.~Section~\ref{sec:results:pbs}.}
    \label{fig:steady_state_estimate}
\end{figure}

\subsubsection{Total Number of Molecules}

We now use \eqref{eq:tf:26} to derive the expected total number of solute and bound particles at time $t$, $N(t)$.
This will allow us to investigate how the enzymatic degradation acts on the solute molecules in the presence of receptor saturation, cf.~Section~\ref{sec:results:deterministic_signal:degradation}.

Now, the number of molecules degraded in interval $[kT,(k+1)T]$ is obtained by subtracting \eqref{eq:tf:26} from \eqref{eq:tf:25}, transforming the result back to the spatial domain similar to \eqref{eq:tf:22}, and integrating output vector $\bm{c}_1$ with respect to $x$.
Assuming that no new molecules are released in $[kT,(k+1)T]$, this yields
\begin{multline}
    N(kT)-N((k+1)T)\\
    = \int_{0}^{a}\bm{c}_1\tran(x)\dint{x}\,(1-\mathrm{e}^{-\kece T})\mathrm{e}^{\As T}\bar{\bm{y}}[k].
\end{multline}
Furthermore, the total number of molecules released by time $(k+1)T$ is given by $\int_{-\infty}^{(k+1)T} s(t) \mathrm{d}t$.
Hence, we obtain
\begin{multline}
    N((k+1)T) = \int_{-\infty}^{(k+1)T} s(t) \mathrm{d}t\\ - \int_{0}^{a}\bm{c}_1\tran(x)\dint{x} \sum_{k'=-\infty}^{k}  (1-\mathrm{e}^{-\kece T})\mathrm{e}^{\As T}\bar{\bm{y}}[k'].\label{eq:surviving_molecules_discrete}
\end{multline}
Choosing $T$ such that $t/T \in \mathbb{Z}$ and setting $k = t/T - 1$, we obtain
\begin{align}
    N((k+1)T) = N\left(\frac{t}{T}T\right) = N(t).\label{eq:surviving_molecules}
\end{align}
We conclude this section noting that the computation of \eqref{eq:surviving_molecules_discrete} comes almost for free when \eqref{eq:tf:26} is evaluated.
In particular, the integral over $x$ can be evaluated offline by direct integration of the trigonometric functions $\cos(\gamma_\mu x)$, cf.~\eqref{eq:tf:6}.

\section{Statistical Signal Model}
\label{sec:statistical_signal_model}
In this section, we first study how the received signal scales with $N$ in the steady state.
Based on the derived nonlinear scaling law, we then derive the signal statistics for the \ac{ISI}-free regime, cf.~\eqref{eq:isi_free_regime}.

Before we start, we setup the notation for this section.
First, we fix some $t=t'$ and introduce the short-hand notations $i=i(t')$ and $I=I(t')$.
Next, we define the binary random variables $M_j \in \lbrace 0,1 \rbrace$ for all $j$, $1 \leq j \leq N$, as
\begin{align}
    M_j = \begin{cases}
        1, \qquad &\textrm{if molecule }j\textrm{ is bound to a receptor at time }t',\\
        0, \qquad &\textrm{otherwise},
    \end{cases}
\end{align}
and denote the probability mass function of $M_j$ by $P_{M_j}(m_j)$, $m_j \in \lbrace 0,1 \rbrace$.
Then, we denote the {\em joint probability distribution} of the sequence of random variables $M_1,\ldots,M_N$ by $P_{M_1,\ldots,M_N}(m_1,\ldots,m_N)$ and the {\em conditional probability distribution} of $M_j$ given $M_1,\ldots,M_{j-1},M_{j+1},\ldots,M_N$ by
\begin{multline}
P_{M_j|M_1,\ldots,M_{j-1},M_{j+1},\ldots,M_N}\\\left(m_j|m_1,\ldots,m_{j-1},m_{j+1},\ldots,m_N\right).
\end{multline}
Finally, we note that $I=\sum_{j=1}^{N} M_j$ and recall that the probability distribution of $I$ is denoted $P_I(n)$, $0 \leq n \leq C^*$.

\subsection{Impact of $N$ and $C^*$ on the Signal Statistics in the Steady State}\label{sec:statistical_signal_model:steady_state}

In this section, we study the impact of the number of molecules, $N$, and the number of receptors, $C^*$, respectively, on the probability that a particle $j$ is bound in the steady state, i.e., as $t \to \infty$.
In this way, we characterize $P_{M_p|M_j}$, $1 \leq p \leq N$, $p \neq j$, in terms of $P_{M_j}$.
This will later allow us to characterize the signal statistics for the transient regime, i.e., when the system is not in steady state.

We consider the instantaneous single release of $N$ molecules in the absence of enzymatic degradation, i.e., $\kece = 0$, $|\mathcal{M}| = 1$.
Furthermore, we assume $t'$ is large enough such that the system is in the steady state.
$i=i^{\infty}$ is then given by Theorem~\ref{thm:steady_state} and $P_{M_j}(1) = 1 - P_{M_j}(0) = i^{\infty}/N$ as all released molecules are identical.

We consider the regime $N \gg C^*$ in which \acp{NT} compete for receptors and, hence, the expected received signal scales nonlinearly in $N$.
Our goal is to characterize the nonlinear dependence of $i$ on $N$.

\begin{lemma}\label{lem:steady_state_approximation}
    Assume $N-j$, $0 \leq j \leq N$, molecules are released, $\kece = 0$, and $N \gg C^*$.
    Further, let $\iinfty(N)$ denote the steady state as given by Theorem~\ref{thm:steady_state} dependent on $N$.
    Then, $\iinfty(N)$ scales as
    \begin{align}
        \iinfty(N-j) \approx \iinfty(N) \frac{N-j}{N-j\left(\iinfty(N)/C^*\right)},\label{eq:lemma_scaling_law}
    \end{align}
    for $0 \leq j \leq N$.
\end{lemma}
\begin{IEEEproof}
    Please see Appendix~\ref{sec:app:proof_lemma_1}.
\end{IEEEproof}
\begin{remark}
    The scaling law proposed in Lemma~\ref{lem:steady_state_approximation} simplifies to $\iinfty(N-j) \approx \iinfty(N) \left(1 - \frac{j}{N}\right)$ if $\iinfty(N)$ is small and to $\iinfty(N-j) \approx \iinfty(N)$ if $\iinfty(N)$ is large.
    This reflects the fact that the impact of $N$ on $\iinfty(N)$ is large if there are relatively few molecules available and that the impact of $N$ on $\iinfty(N)$ is negligible if molecules are abundant.
\end{remark}

We investigate the accuracy of \eqref{eq:lemma_scaling_law} in Figure~\ref{fig:steady_state_estimate}.
In particular, Figure~\ref{fig:steady_state_estimate} shows that \eqref{eq:lemma_scaling_law} captures the nonlinear dependence of $\iinfty$ on $N$.

We now return to our initial goal of computing $P_{M_p|M_j}(m_p|m_j)$.
To this end, we consider first $P_{M_p|M_j}(1|0)$, i.e., the probability that molecule $p$ is bound given that molecule $j$ is not bound.
As the binding of molecules to receptors is competitive, the probability that one of the remaining $N-1$ molecules binds to a receptor increases, if $j$ is not bound.
As all of the $N-1$ molecules are identical, $P_{M_p|M_j}(1|0)$ is then given as the ratio of the expected number of bound molecules when only $N-1$ molecules are released and the total number of molecules $N-1$, i.e.,
\begin{align}
    P_{M_p|M_j}(1|0) = \frac{\iinfty(N-1)}{N-1}.
\end{align}

Exploiting Lemma~\ref{lem:steady_state_approximation}, we obtain the following model
\begin{align}
    P_{M_p|M_j}(1|0) \approx \frac{\iinfty(N)}{N-\iinfty(N)/C^*}.
\end{align}

Now, we consider $P_{M_p|M_j}(1|1)$.
When molecule $j$ is bound, the probability that one of the remaining $N-1$ molecules binds to a receptor decreases.
Similar to the scaling law presented in Lemma~\ref{lem:steady_state_approximation}, we assume that the effect of reducing the number of available receptors can be modeled by a scaling factor, i.e.,
\begin{align}
    P_{M_p|M_j}(1|1) \approx \alpha \frac{i}{N-i/C^*},
\end{align}
where $\alpha$ is the unknown scaling factor yet to be determined, and in slight abuse of the short-hand notation introduced at the beginning of this section, we write $i=\iinfty(N)$ .

To determine $\alpha$, we recall that the marginal binding probabilities are equal for all particles.
Marginalizing over $M_j$, we therefore obtain
\begin{align}
    i/N &= P_{M_p}(1)\nonumber\\
    &= P_{M_p|M_j}(1|0)P_{M_j}(0) + P_{M_p|M_j}(1|1)P_{M_j}(1)\nonumber\\
    &\approx \frac{i}{N-i/C^*} \left(1 - \frac{i}{N}\right) + \alpha \frac{i}{N-i/C^*} \frac{i}{N}.\label{eq:alpha_marginal}
\end{align}
Solving \eqref{eq:alpha_marginal} for $\alpha$ yields $\alpha=1-1/C^*$.

In summary, we obtain
\begin{align}
    P_{M_p|M_j}(1|m_j) &= 1 - P_{M_p|M_j}(0|m_j)\nonumber\\ &\approx \begin{cases}
        \frac{N-1}{N-i/C}\frac{i}{N-1} & \, \textrm{ if } m_j = 0 \\
        \frac{C^*-1}{C^*}\frac{N-1}{N-i/C^*}\frac{i}{N-1} & \, \textrm{ if } m_j = 1
    \end{cases}.
\end{align}

Hence, the impact of $N$ on $P_{M_p|M_j}$ is characterized by the term $(N-1)/(N-i/C^*)$, while the impact of $C^*$ is characterized by $(C^*-1)/C^*$.
We now generalize this model to $P_{M_p|M_{p-1},\ldots,M_{1}}$ for $1 \leq p \leq N$ as follows
\begin{IEEEeqnarray}{rCl}
    \IEEEeqnarraymulticol{3}{l}{
    P_{M_p|M_{p-1},\ldots,M_{1}}(1|m_{p-1},\ldots,m_1)}\nonumber\\
        &=& 1 - P_{M_p|M_{p-1},\ldots,M_{1}}(0|m_{p-1},\ldots,m_1)\nonumber\\
        &\approx& \frac{C^*-\sum_{j=1}^{p-1}m_j}{C^*}\frac{N-p+1}{N-(p-1)(i/C^*)}\frac{i}{N-p+1},\label{eq:statistical_model_steady_state}
\end{IEEEeqnarray}
where the linear term $(C^*-\sum_{j=1}^{p-1}m_j)/C^*$ models the reduction of the binding probability due to the binding of molecules $1,\ldots,p-1$.
In particular, we note that if $\sum_{j=1}^{p-1}m_j = C^*$, the binding probability for molecule $p$ equals $0$.
The nonlinear term $(N-p+1)/[N-(p-1)(i/C^*)]$ results from Lemma~\ref{lem:steady_state_approximation}.

In the next subsection, we generalize this model further to the statistics of the received signal at any time $t$ in the presence of enzymatic degradation.

\subsection{Signal Statistics for \ac{ISI}-free Scenario}\label{sec:statistical_signal_model:single_release}

We recall from the discussion in Section~\ref{sec:system_model:system_model:statistics} that the statistics of $I(t)$ approximately follow the binomial distribution $B(n;i(t)/C^*,C^*)$ if the receptors operate independently, i.e., if many \acp{NT} are present relative to receptors.
This is the case, if the number of released \acp{NT} or the number of residual \acp{NT} due to \ac{ISI} is large.
However, if \ac{ISI} is negligible, i.e., \eqref{eq:isi_free_regime} is fulfilled, the competition of receptors for \acp{NT} can become significant and the binomial model does not apply.
Hence, in this section, we consider the statistics of the received signal under the assumption that \ac{ISI} is negligible.
Under this assumption, it is sufficient to consider the single instantaneous release of $N$ molecules at $t=0$.
As in the former subsection, we write $i = i(t')$ and $I = I(t')$.

As all $N$ molecules are identical, we have
\begin{align}
    P_{I}(n) &= \sum_{\substack{
            m_1,\ldots,m_N \in \lbrace 0,1 \rbrace^N\\
            m_1 + \ldots + m_N = n
    }} P_{M_1,\ldots,M_N}(m_1,\ldots,m_N)\nonumber\\
    &= \binom{N}{n} P_{M_1,\ldots,M_n,M_{n+1},\ldots,M_N}(1,\ldots,1,0,\ldots,0).\label{eq:statistical_model}
\end{align}

This observation leads to the following characterization of $P_I$.

\begin{theorem}\label{thm:PI_PH}
    Assume $NC^*/i$ is an integer, $N \gg C^*$, and $i \leq \frac{C^*}{1+C^*/N}$.
    Then,
    \begin{align}
        P_{I}(n) \approx P_{H}(n;NC^*/i,C^*,N),\label{eq:statistical_model_hgeom}
    \end{align}
    where $P_{H}(n;r'+b',r',k')$ denotes the {\em hypergeometric distribution} with parameters $r'+b',r',k'$, $0 \leq n \leq r$, $0 \leq k'-n \leq b'$, $1 \leq k' \leq r'+b'$,
    \begin{align}
        P_{H}(n;r'+b',r',k') = \frac{\binom{r'}{n}\binom{b'}{k'-n}}{\binom{r'+b'}{k'}}.
    \end{align}
\end{theorem}
\begin{IEEEproof}
    Please see Appendix~\ref{sec:app:proof_thm_2}.
\end{IEEEproof}
\begin{remark}
    The hypergeometric distribution models {\em drawing without replacement}.
    As such, it is a simple model for {\em negative dependence} which is encountered in the context of statistical physics.
    There, it is used to model different physical and biological phenomena in which the action of one particle or molecule prevents some action by another molecule.
    Examples include the energy state occupancy by fermions \cite{dubhashi96} and the movement of molecules towards free vertices along the edges of a graph \cite{pemantle00}.
    Hence, it is not surprising that we obtain the hypergeometric distribution for the problem at hand.
\end{remark}
\begin{remark}
    Negative dependence between random variables is closely associated with the set of {\em ultra-log-concave functions} \cite{pemantle00}.
    Within this class of functions, the binomial distribution is the limiting distribution if the statistical dependence between events approaches $0$ and maximizes the entropy \cite{yu08}.
    Hence, assuming binomial noise when the actual noise distribution follows the hypergeometric model leads to a potential overestimation of the counting noise.
\end{remark}
\begin{remark}
    If $C^*$ grows large, $P_{H}(n;NC^*/i,C^*,N)$ approaches the binomial distribution $\mathcal{B}(i/N,N)$.
    This case corresponds to the case in which the statistical dependence between the molecules is negligible.
\end{remark}
\begin{remark}
    If $N$ grows large but $C^*$ remains constant, such that $i/C^*$ tends towards $1$, the variance of $P_{H}(n;NC^*/i,C^*,N)$ tends towards $0$.
    Hence, by releasing many molecules, almost deterministic transmission can by achieved (on the expense, however, of large \ac{ISI}).
\end{remark}
\begin{remark}
    $P_{H}(n;NC^*/i,C^*,N)$ is symmetric in the second and the third parameter, i.e.,
    \begin{align}
        P_{H}(n;NC^*/i,C^*,N)=P_{H}(n;NC^*/i,N,C^*).
    \end{align}
    While we have derived $P_{H}(n;NC^*/i,C^*,N)$ for the regime in which molecules compete for receptors, i.e., $N \gg C^*$, this symmetry suggests that $P_{H}(n;NC^*/i,C^*,N)$ is also a valid model for the competition of receptors for molecules.
    Indeed, assuming $C^* \gg N$, we can use a similar line of argumentation as in Section~\ref{sec:statistical_signal_model:steady_state} to arrive at $P_{H}(n;NC^*/i,N,C^*)$.
    As this involves basically the same steps as presented in Section~\ref{sec:statistical_signal_model:steady_state}, we omit the derivation here due to space constraints.
    However, the regime $C^* \gg N$ is further discussed in Section~\ref{sec:results}.
\end{remark}
\begin{remark}
    Finally, as a direct consequence of the previous remark, $P_{H}(n;NC^*/i,C^*,N)$ approaches $\mathcal{B}(i/C^*,C^*)$ if $N$ grows large.
\end{remark}

\begin{corollary}
    The mean and the variance of the number of bound particles under the statistical model in \eqref{eq:statistical_model_hgeom} are given as
    \begin{equation}
        \mathrm{E}_{H}\lbrace I \rbrace = i,\label{eq:PH_mean}
    \end{equation}
    and
    \begin{equation}
        \textrm{Var}(I) = \mathrm{E}_{H}\lbrace (I-\mathrm{E}_{H}\lbrace I \rbrace)^2 \rbrace = i \frac{(1-i/N) (1-i/C^*)}{1-i/(NC^*)},\label{eq:PH_var}
    \end{equation}
    respectively, where $\mathrm{E}_{H}$ denotes expectation \ac{wrt} the distribution defined in \eqref{eq:statistical_model_hgeom}.
\end{corollary}
\begin{IEEEproof}
    Eqs.~\eqref{eq:PH_mean} and \eqref{eq:PH_var} follow directly from the mean and the variance, respectively, of the hypergeometric distribution.
\end{IEEEproof}

\section{Numerical Results}
\label{sec:results}
In this section, we study the deterministic and statistical signal models derived in Sections~\ref{sec:deterministic_signal_model} and \ref{sec:statistical_signal_model} for some exemplary sets of parameter values.
Furthermore, we present the results from three-dimensional \ac{PBS} to verify the accuracy of the analytical models and approximations presented in Sections~\ref{sec:system_model}, \ref{sec:deterministic_signal_model}, and \ref{sec:statistical_signal_model}.

\subsection{Particle-based Simulation and Choice of Parameters}\label{sec:results:pbs}

The basic design of the particle simulator was adopted from \cite{lotter20} and we refer the interested reader to \cite{lotter20} for further details.
Receptor saturation was incorporated into the simulator presented in \cite{lotter20} by setting the binding probability for a receptor to zero when a molecule was bound to this receptor, and back to its original value when the molecule unbound.
Enzymatic degradation was incorporated by introducing a first-order degradation step for all solute molecules with probability \cite{andrews09} $1 - \exp(-\kappa_e C_E \Delta t)$, where $\Delta t$ denotes the simulation time step in \si{\micro\second}.
Furthermore, the scaling parameter for the boundary homogenization from \cite{lotter20} was slightly adapted to reflect the larger receptor radius used in this paper compared to \cite{lotter20}.
To compute the modified scaling parameter, the same method as in \cite{lotter20} was used, i.e., the steady state number of bound molecules was simulated for different parameter values in the absence of enzymatic degradation and then compared to \eqref{eq:i_infty} to fit $\ka$.
From this procedure, we obtained $\ka = 0.995 \rho \kad$, where $\rho$ denotes the fraction of the postsynaptic membrane occupied by receptors.
If not indicated otherwise, the results from \ac{PBS} were averaged over $150$ realizations.

The computational cost of the \ac{PBS} scales with the simulation time step as well as with the number of released particles, the number of receptors, and the number of simulation runs.
The runtime of the proposed \ac{SSD} model, in contrast, scales only with the sampling interval and the number of eigenfunctions.
Consequently, for the parameter values considered in this paper, the computation of the \ac{SSD} model required far less (by more than a factor of $100$) CPU time than the \ac{PBS}.

If not indicated otherwise, the default parameter values presented in Table~\ref{tab:sim_params} were used.
\begin{table}
    \centering
    \caption{Simulation parameters for particle-based simulation \cite{lotter20}.}
    \footnotesize
    \begin{tabular}{| p{.16\linewidth} | r | p{.38\linewidth} |}
        \hline Parameter & Default Value & Description\\ \hline
        $D$ & $\SI{3.3e-4}{\micro\meter\squared\per\micro\second}$& Diffusion coefficient\\ \hline
        $N$ & $\SI{1000}{}$ & Number of released particles\\ \hline
        $a$ & $2 \times 10^{-2}$~$\si{\micro\meter}$& Channel width in $x$\\ \hline
        $\{y,z\}_{\textrm{max}}-\{y,z\}_{\textrm{min}}$ & $\SI{0.15}{\micro\meter}$ & Channel widths in $y$ and $z$\\ \hline
        $\kad$ &$1.02 \times 10^{-4}$~$\si{\micro\meter\per\micro\second}$ & Intrinsic binding rate\\ \hline
        $\kd$ &$8.5 \times 10^{-3}$~$\si{\per\micro\second}$ & Intrinsic unbinding rate\\ \hline
        $\kappa_e C_{E}$ &$10^{-3}$~$\si{\per\micro\second}$ & Degradation rate\\ \hline
        $r$ & $2.3 \times 10^{-3}$~$\si{\micro\meter}$ & Receptor radius\\ \hline
        $C^*$ & $203$ & Number of uniformly distribu\-ted receptors (15\% coverage)\\ \hline
        $\Delta t$ & $10^{-2}$~$\si{\micro\second}$ & Simulation time step\\ \hline
        $Q$ & $100$ & Number of eigenfunctions\\ \hline
        $T$ & $1 \times 10^{-1}$~$\si{\micro\second}$ & Sampling interval\\ \hline
    \end{tabular}
    \label{tab:sim_params}
\end{table}

\subsection{Expected Received Signal}

\subsubsection{Saturation and Enzymatic Degradation for Single Release}

\begin{figure}[t]
    \includegraphics[width=\linewidth]{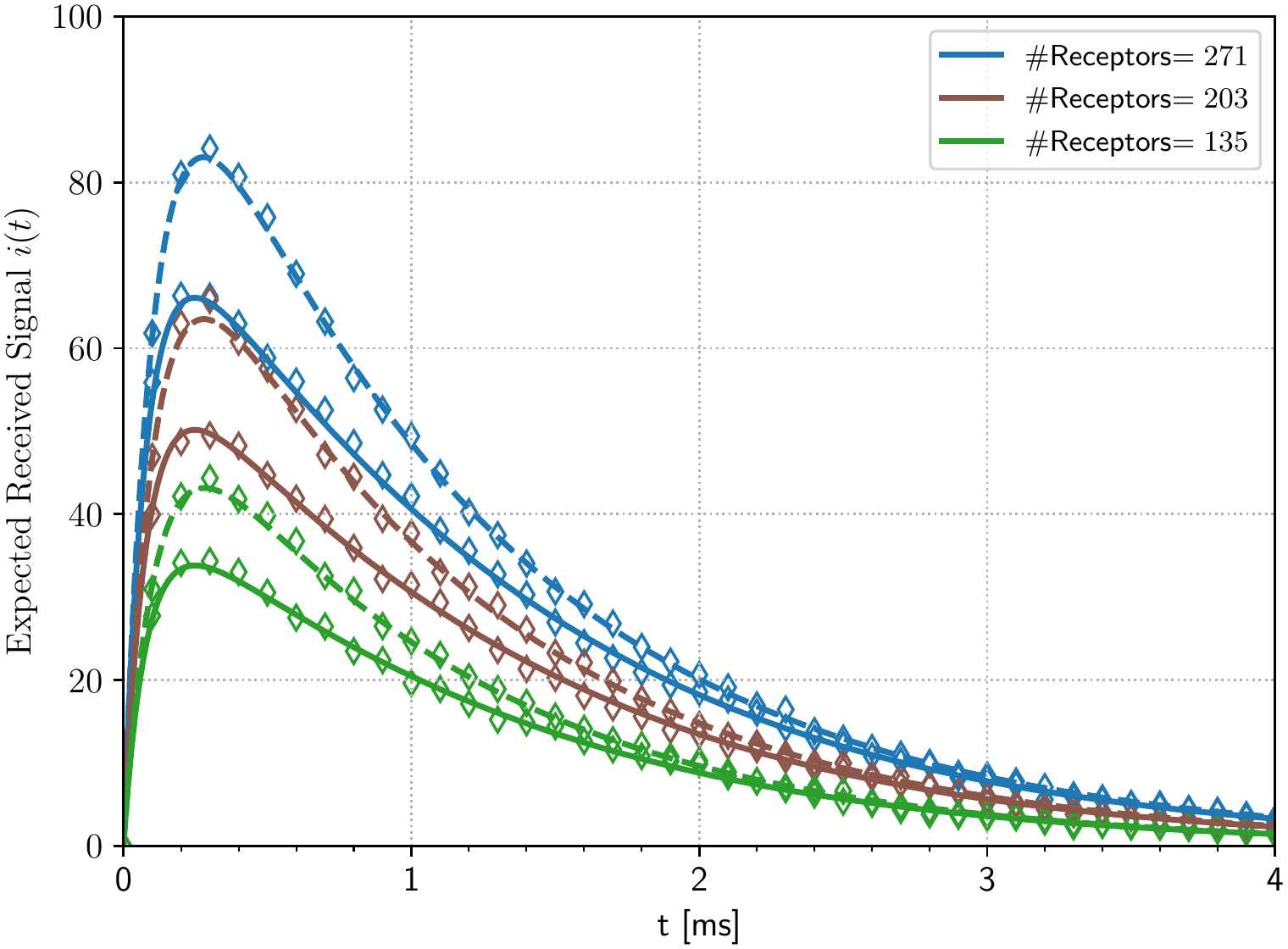}
    \caption{Expected received signal according to \eqref{eq:tf:13} with (solid lines) and without (dashed lines) saturation for different numbers of postsynaptic receptors in the presence of enzymes. Results from \ac{PBS} are shown as diamond markers.}
    \label{fig:det_single_release}
\end{figure}

In this section, the impact of saturation on the expected received signal $i(t)$ in the presence of enzymatic degradation is investigated.
Fig.~\ref{fig:det_single_release} shows $i(t)$ as computed with the \ac{SSD} defined by \eqref{eq:tf:26} and \eqref{eq:tf:13} for different numbers of postsynaptic receptors with and without saturation.
It can be observed in Fig.~\ref{fig:det_single_release} that saturation reduces the peak value as compared to the system without saturation.
Furthermore, we observe that the peak value depends approximately linearly on the number of available receptors $C^*$ which confirms the approximation made in Theorem~\ref{thm:PI_PH}.

\subsubsection{The Impact of Saturation on Multiple Releases}

\begin{figure}[t]
    \includegraphics[width=\linewidth]{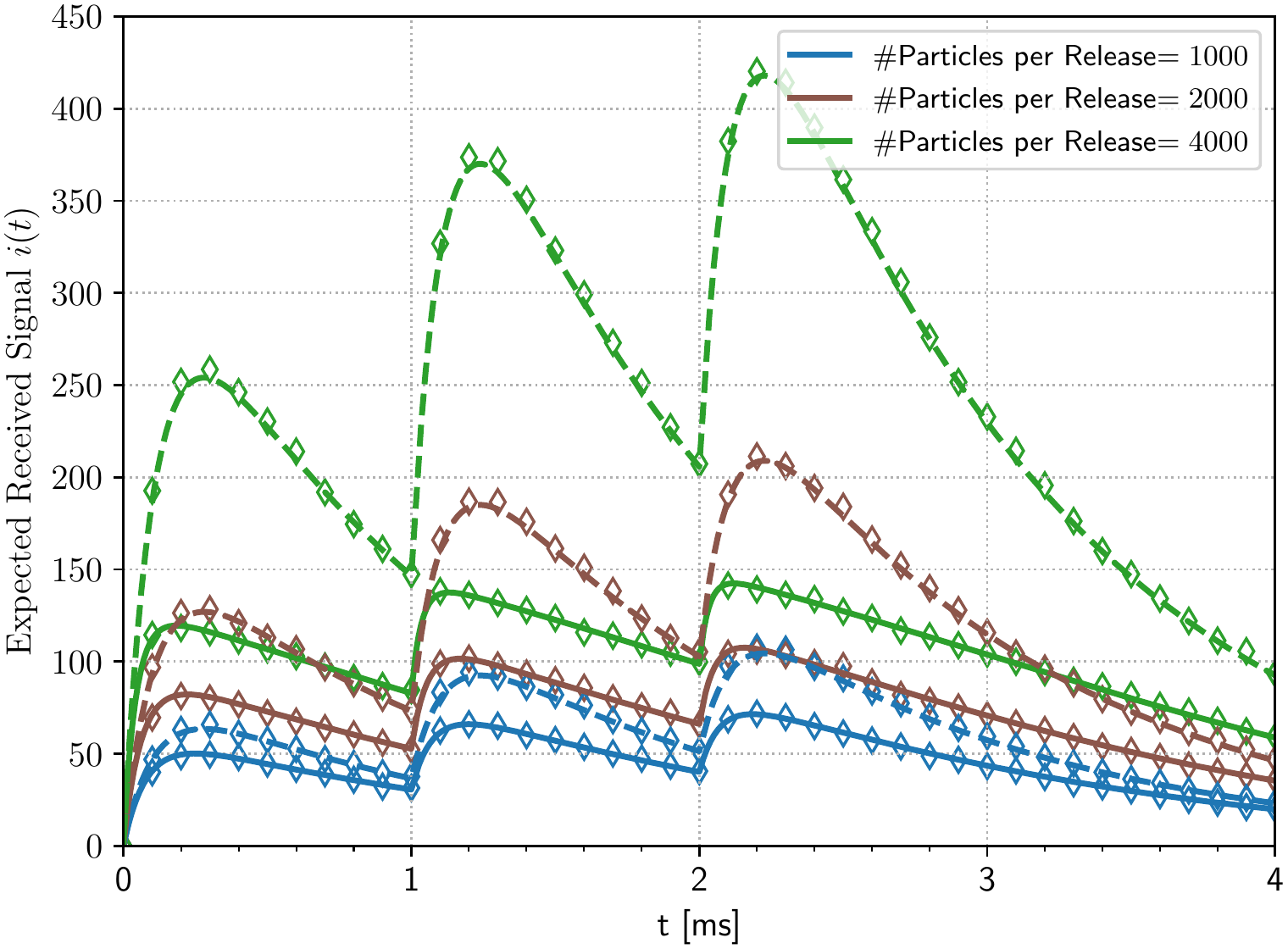}
    \caption{Results from \ac{PBS} (diamond markers) and \eqref{eq:tf:13} (lines) for different numbers of released molecules $N$ with (solid lines) and without (dashed lines) saturation when enzymes are present.}
    \label{fig:det_multiple_releases}
\end{figure}

Next, we investigate the impact of receptor saturation on the postsynaptic signal for multiple \ac{NT} releases when enzymes are present.
Molecules were released at $T_0 = \SI{0}{\micro\second}$, $T_1 = \SI{1}{\micro\second}$, and $T_2 = \SI{2}{\micro\second}$.
The results for the \ac{SSD} presented in Section~\ref{sec:deterministic_signal_model:state-space_description} and \ac{PBS} are shown in Fig.~\ref{fig:det_multiple_releases} for different numbers of released molecules $N$.
First, we observe that in the presence of receptor saturation, the peaks of the postsynaptic signals do not scale linearly with $N$.
Next, we observe that the impact of receptor saturation in terms of the peak values becomes more pronounced compared to the system without saturation as $N$ increases.
Finally, we note that, due to \ac{ISI}, for each $N$ the peak value following the second release of \acp{NT} (at $t \approx \SI{1.2}{\milli\second}$) is larger than the peak value following the first release.
Now, interestingly, this effect is significantly less pronounced in the presence of saturation.
In fact, this observation is consistent with experimental observations \cite{foster05} and has two reasons.
First, limiting the number of receptors naturally damps the signal because fewer receptors are available.
Second, in the presence of receptor saturation, fewer molecules are bound simultaneously as compared to the case without receptor saturation and, consequently, molecules become more exposed to degradation and the channel is cleared faster\footnote{The last conclusion requires the first-order assumption from A\ref{ass:enz_deg}, cf.~Section~\ref{sec:system_model:assumptions}.}.

\subsubsection{The Impact of Receptor Saturation on Enzymatic Degradation}\label{sec:results:deterministic_signal:degradation}

From all molecules present at time $t$ as either solute molecules or bound molecules, only the solute molecules are exposed to enzymatic degradation.
Hence, due to the ``buffering'' of molecules at postsynaptic receptors, the overall degradation rate of \acp{NT} is slower than $\kece$.
The competition of molecules for receptors, on the other hand, limits the number of concurrently bound molecules as compared to the case without receptor saturation, i.e., the impact of buffering is reduced.

\begin{figure}[t]
    \includegraphics[width=\linewidth]{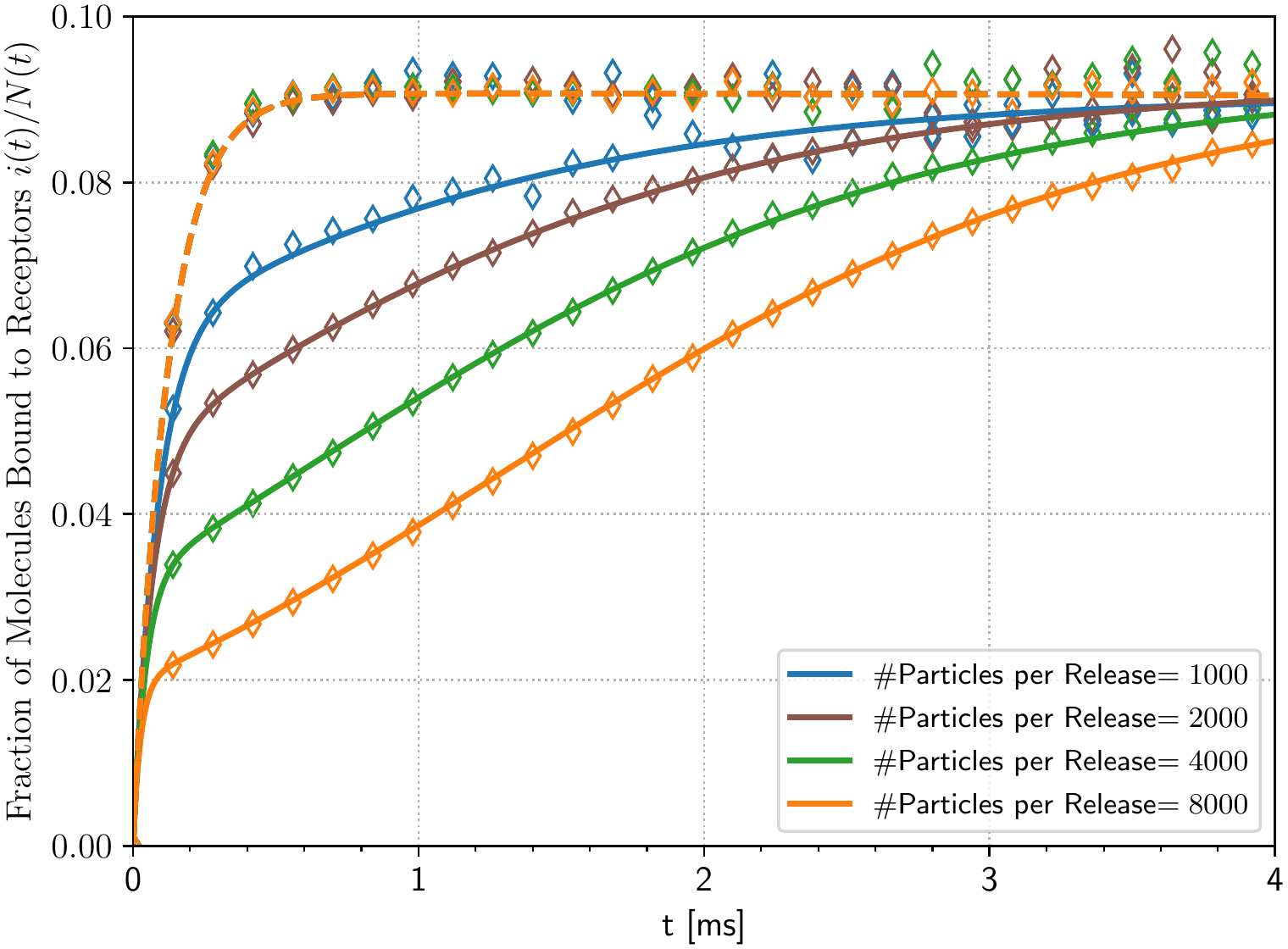}
    \caption{Fraction of molecules bound to receptors at time $t$ in the presence (solid lines) and absence (dashed lines) of saturation. The curves for different values of $N$ coincide in the absence of saturation. The results from \ac{PBS} are shown as diamond markers.}
    \label{fig:degradation}
\end{figure}

In this section, we investigate the impact of receptor saturation on the ratio of bound molecules to the total number of particles at time $t$, $i(t)/N(t)$, where the expected total number of solute and bound particles at time $t$, $N(t)$, is given by \eqref{eq:surviving_molecules}.
Fig.~\ref{fig:degradation} shows $i(t)/N(t)$ after a single release of $N$ molecules for different $N$ with and without saturation.
We observe from Fig.~\ref{fig:degradation} that $i(t)/N(t)$ does {\em not} depend on $N$ if receptor saturation is neglected.
In this case, $i(t)/N(t)$ rises initially and then remains approximately constant after $t \approx 400$~$\si{\micro\second}$.
In the presence of receptor saturation, on the other hand, we observe from Fig.~\ref{fig:degradation} that $i(t)/N(t)$ is an increasing function of $t$ in the considered time frame.
Hence, we conclude that $N(t)$ decays at a faster rate than $i(t)$ in the presence of receptor saturation, while $N(t)$ and $i(t)$ decay approximately with the same rate in the absence of saturation.
Furthermore, in the presence of saturation, $i(t)/N(t)$ depends on $N$.
In particular, as $N$ is increased, the initial steep rise of $i(t)/N(t)$ in the interval $t \in [\SI{0}{\micro\second}; \SI{200}{\micro\second}]$ is decreased.
This effect is indeed expected, because the total number of molecules scales initially linearly with $N$, while $i(t)$ is bound by the number of available receptors $C^*$.

Finally, we observe in Fig.~\ref{fig:degradation} that the variance of the \ac{PBS} data increases as $t$ increases. This increase in variance is due to the small number of molecules $N(t)$ for large $t$ and underlines the usefulness of the proposed deterministic model. Namely, while the expected system state could be obtained by further extensive averaging of \ac{PBS} data, this would incur considerable computational cost. In contrast, the proposed \ac{SSD} model readily yields the expected system state without diffusion noise for all $t$, irrespective of the number of molecules present.

\begin{figure}[t]
    \includegraphics[width=\linewidth]{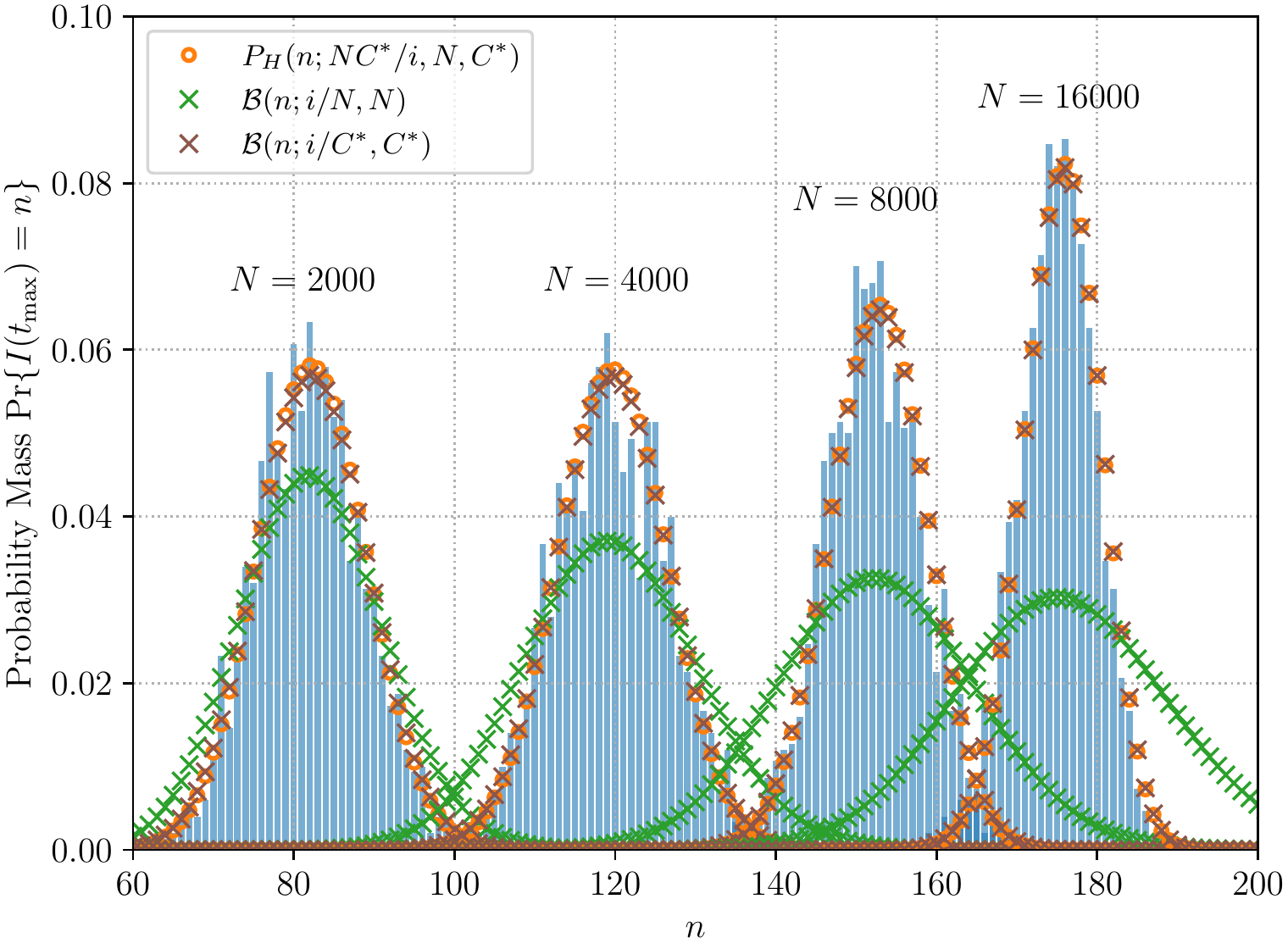}
    \caption{Single release statistics at $\tmax$ for different values of $N$. The empirical distribution obtained with \ac{PBS} is shown in blue. The hypergeometric model as proposed in Section~\ref{sec:statistical_signal_model:single_release} is shown as orange circles and the binomial models as discussed in Section~\ref{sec:system_model:system_model:statistics} are shown in green and brown for comparison.}
    \label{fig:stats_single_release_N}
\end{figure}

\subsection{Statistics of the Received Signal}

In this section, we investigate the statistics of the received signal for one single release of $N$ molecules.
To this end, we computed $6000$ random realizations of the \ac{PBS} for each considered parameter set to obtain the empirical distributions of the received signal at different time instants $t$.

\subsubsection{Competition of Molecules for Receptors}

First, we consider the release of $N$ particles, where $N \gg C^*$, and investigate the statistics of $I(t_{\mathrm{max}})$, where $I(\tmax)$ denotes the random number of bound receptors at the expected time of peak, i.e., $\tmax = \operatorname{arg\,max}_t i(t)$.
Fig.~\ref{fig:stats_single_release_N} shows the empirical distributions of $I(t_{\mathrm{max}})$ for different $N$ and the statistical model proposed in Section~\ref{sec:statistical_signal_model:single_release}.
We observe from Fig.~\ref{fig:stats_single_release_N} that the agreement between the proposed statistical model and the empirical data is very good in all considered regimes.
Further, we observe that the variance of $I(\tmax)$ decreases as $N$ increases towards $N=16,000$.
This effect is indeed expected intuitively because a larger number of molecules effectively averages out the diffusion noise.
As can be observed in Fig.~\ref{fig:stats_single_release_N}, the binomial model $\mathcal{B}(i/N,N)$ is not able to capture the impact of saturation on the signal statistics.
This is due to the fact that in the considered parameter regime the molecules compete for receptors and, hence, the statistical independence assumption underlying the binomial distribution $\mathcal{B}(i/N,N)$ is not fulfilled.
On the other hand, the binomial model $\mathcal{B}(i/C^*,C^*)$ provides a good approximation of the signal statistics, as the competition of receptors for molecules is not relevant in the considered parameter regime and, hence, the receptors are approximately statistically independent.

\subsubsection{Competition of Receptors for Molecules}

\begin{figure}[t]
    \includegraphics[width=\linewidth]{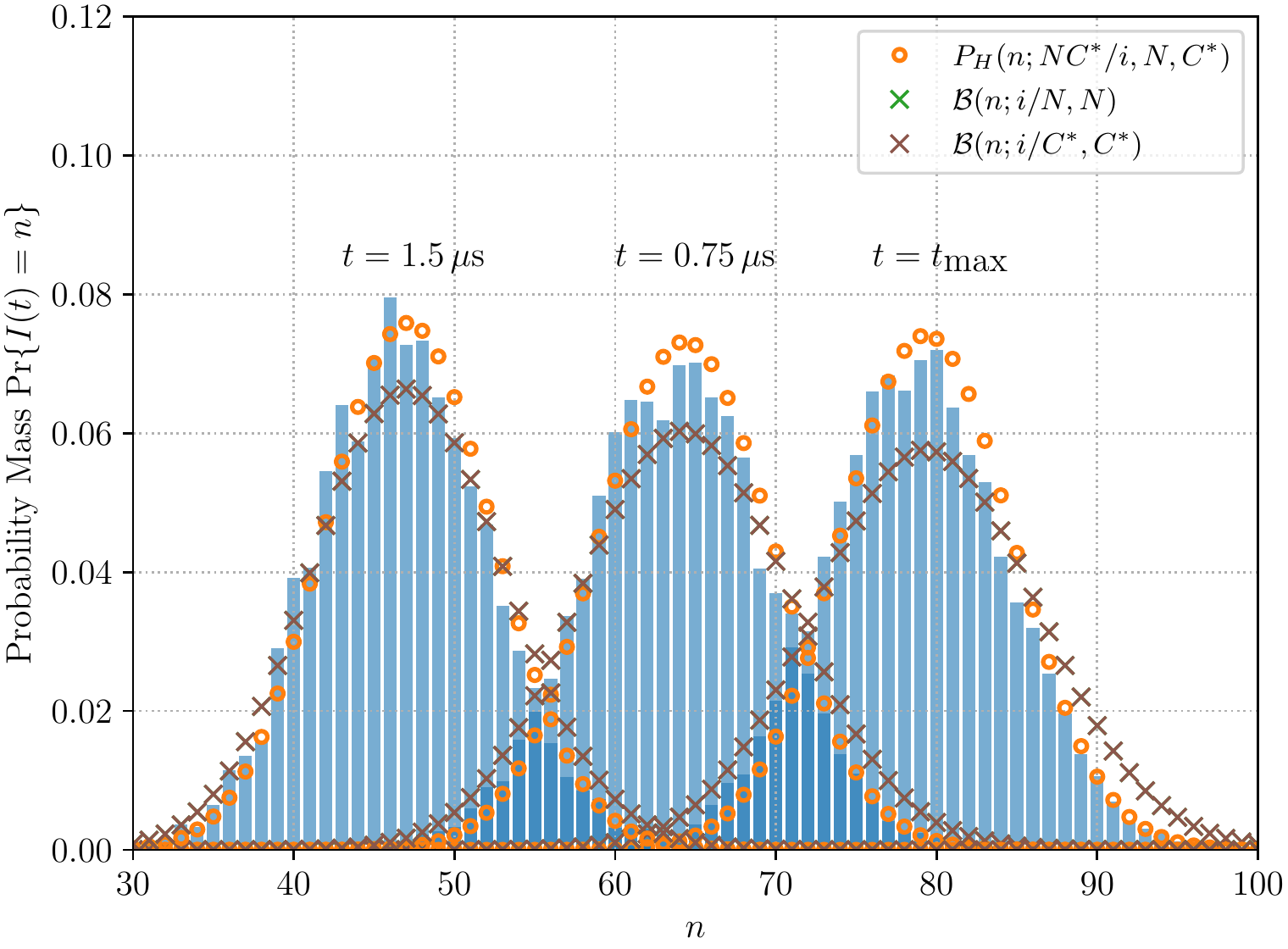}
    \caption{Single release statistics at different times $t$. The empirical distribution obtained with \ac{PBS} is shown in blue. The hypergeometric model as proposed in Section~\ref{sec:statistical_signal_model:single_release} is shown as orange circles and the binomial models as discussed in Section~\ref{sec:system_model:system_model:statistics} are shown in green and brown (indistinguishable in figure) for comparison.}
    \label{fig:stats_single_release_t}
\end{figure}

Next, we consider the signal statistics at different time instants $t \in \lbrace \tmax, \SI{0.75}{\micro\second}, \SI{1.5}{\micro\second} \rbrace$ for a parameter regime in which receptors compete for molecules.
Namely, we set the receptor radius to $r = 4 \times 10^{-3}\,\si{\micro\meter}$, the intrinsic binding rate of the receptors to $\kad = 5.14 \times 10^{-4}\,\si{\micro\meter\per\micro\second}$, and the number of released particles to $N=200$.
This parameter regime models a scenario with increased competition of receptors for molecules as compared to the reference scenario defined in Table~\ref{tab:sim_params}.
The competition is due to the fact that fewer molecules ($N$) are present, these molecules are more likely to hit a receptor (because of the larger $r$) and remain trapped at the receptor for a longer time (because of the larger $\kad$).
From Fig.~\ref{fig:stats_single_release_t}, we observe that the proposed statistical model matches the data obtained by \ac{PBS} very well.
Furthermore, Fig.~\ref{fig:stats_single_release_t} shows that both binomial models, $\mathcal{B}(i/N,N)$ and $\mathcal{B}(i/C^*,C^*)$, deviate from the observed empirical distribution.
In particular, the actual distribution of the received signal is more concentrated than predicted by the binomial models.
This deviation is a consequence of the independence assumption underlying the binomial distribution which is not fulfilled in the presence of competition.
Specifically, $\mathcal{B}(i/C^*,C^*)$ does not provide an accurate model for the signal statistics in Fig.~\ref{fig:stats_single_release_t}, because in the scenario considered in Fig.~\ref{fig:stats_single_release_t} the probability that a given receptor is activated by a solute \ac{NT} at time $t$ depends on the actual number of bound \acp{NT} at that time instant, i.e., in contrast to the scenario considered in Fig.~\ref{fig:stats_single_release_N}, the receptors cannot be assumed to act independently.

The present example shows that the statistical model proposed in Section~\ref{sec:statistical_signal_model:single_release} provides an accurate model for the statistics of the received signal even in the presence of receptor competition while the binomial models are not able to capture this effect.

\subsubsection{The Impact of System Parameters on the Signal Variance}

\begin{figure}[t]
    \includegraphics[width=\linewidth]{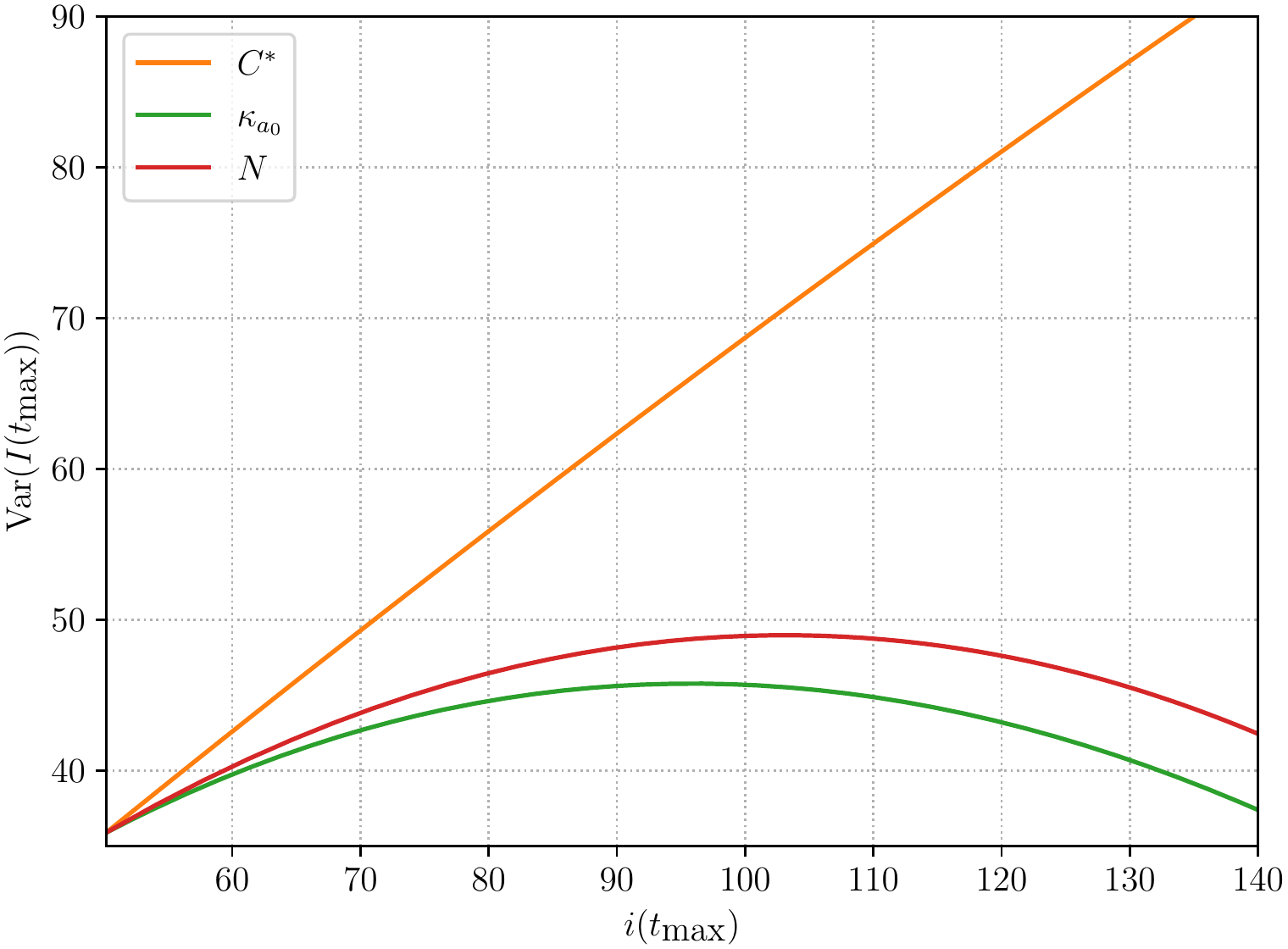}
    \caption{Variance of the received signal at the peak of the expected received signal as the system parameters are varied.}
    \label{fig:variance_at_peak}
\end{figure}

In this section, we study the impact of the various system parameters on the signal variance given by Eq.~\eqref{eq:PH_var}.
In particular, we compare the effects of
\begin{itemize}
    \item increasing the number of receptors $C^*$,
    \item increasing the number of released molecules $N$, and
    \item increasing the intrinsic binding rate of the receptors $\kad$
\end{itemize}
on the variance of the received signal at its peak value after $N$ particles have been released.
Clearly, all the mentioned adjustments lead to a larger peak value $i(\tmax)$ of the expected signal.
However, as Fig.~\ref{fig:variance_at_peak} shows, their respective impacts on the signal variance are very different.

First, we observe from Fig.~\ref{fig:variance_at_peak} that increasing $C^*$ leads to a linear increase in the variance.
In contrast, if $\kad$ or $N$ are increased, the variance peaks at an intermediate value of $i$ and decreases towards larger values of $i$.
This observation is intuitive as the signal variance ultimately depends on the ratio between the available molecules and the available receptors.
We note that the trade-off between $i$ and $\textrm{Var}(I)$ that we observe in Fig.~\ref{fig:variance_at_peak} is specific to the saturating receiver and the presented example shows that the proposed model can deliver novel insights which may prove useful with respect to system design.
In particular, Fig.~\ref{fig:variance_at_peak} reveals the impact of $N$, $C^*$, and $\kad$ on the signal-dependent noise which may have implications for transmitter ($N$) and receiver design ($C^*$ and $\kad$).

\section{Conclusion}
\label{sec:conclusion}
In this paper, we presented novel deterministic and statistical signal models for synaptic \ac{DMC} in the presence of enzymatic degradation.
The proposed deterministic model is based on a nonhomogeneous reaction-diffusion equation with nonlinear boundary condition and is solved using an \ac{SSD} obtained from a \ac{TFM} formulation of the boundary-value problem.
The resulting \ac{SSD} provides an efficient means to investigate the (relative) impact of the various biophysical mechanisms of synaptic \ac{DMC} on the received signal.
It allows, for example, to evaluate how the expected number of bound receptors scales with the number of available receptors, the postsynaptic binding kinetics, the enzymatic degradation rate, and the number of released molecules, respectively, in the presence of receptor saturation.
As an exemplary use case, we have investigated how the number of postsynaptic receptors and the number of \acp{NT} per release shape the expected received signal.
Results from \ac{PBS} have confirmed that the proposed model captures the expected received signal accurately.

Furthermore, a novel statistical model in terms of the hypergeometric distribution is proposed for the \ac{ISI}-free regime.
The proposed statistical model is derived from the proposed deterministic model by analyzing the dependence of the steady state number of bound receptors on the number of released molecules and making suitable approximations.
While existing models rely on the assumption of statistical independence either between molecules or between receptors, the statistical model proposed in this paper captures the impact of the competition of \acp{NT} for receptors {\em and} the impact of the competition of receptors for \acp{NT} on the signal statistics.
Since both types of competition can be relevant in nature \cite{zucker14}, we expect the proposed model to be helpful for enhancing the current understanding of the operational challenges and limits of chemical synapses.
The accuracy of the proposed statistical model is verified by \ac{PBS}.

In summary, we envision the proposed framework to contribute to the understanding of the chemical synapse as one of the building blocks of neuronal communication.
This may enable the design of sophisticated synthetic \ac{DMC} systems.
A potential target for future work could be the extension of the proposed framework to account for synaptic clearance mechanisms other than enzymatic degradation.
For example, the uptake of \acp{NT} at glial cells or presynaptic reuptake \cite{zucker14} could be considered.
As further extension of the proposed models, it would be interesting to see how they can be extended to other MC systems, e.g.~systems employing the widely considered spherical receiver.

\appendix
\section{}
\subsection{Proof of Theorem~\ref{thm:steady_state}}\label{sec:app:proof_thm_1}
In the steady state, there is no net flux of molecules across the boundaries $x=0$ and $x=a$, i.e., $\lim\limits_{t \to \infty} i_x(0,t) = \lim\limits_{t \to \infty} i_x(a,t) = 0$.
Furthermore, because $c(x,t)$ is twice differentiable \ac{wrt} $x$ and its first derivative \ac{wrt} $x$ at both boundaries vanishes in the steady state, $c(x,t)$ is constant in $x$ for $t \to \infty$, i.e., $\lim\limits_{t \to \infty} c(x,t) = c^{\infty},\, \forall x \in (0;a)$.
Finally, we have the following conservation equation
\begin{align}
    \int_{0}^{a} c^{\infty} \mathrm{d}x + i^{\infty} = a c^{\infty} + i^{\infty} = \int_{-\infty}^{\infty} s(\tau) \mathrm{d}\tau = N |\mathcal{M}|,\label{eq:conservation_equation}
\end{align}
because $\kece=0$ and, hence, no molecules are lost due to enzymatic degradation.
Plugging \eqref{eq:conservation_equation} into \eqref{eq:pab_rb} and equating the right-hand side of \eqref{eq:pab_rb} to $0$, we obtain \eqref{eq:steady_state}
It is left to show that the smallest root of \eqref{eq:steady_state} is the relevant one.
To this end, we observe that
\begin{IEEEeqnarray}{rCl}
    \IEEEeqnarraymulticol{3}{l}{
    \left[\left(1+\frac{a \kd}{\ka}\right)C^* + N |\mathcal{M}|\right]^2 - 4 N |\mathcal{M}| C^*}\nonumber\\
    &\geq& \left(C^*\right)^2 + 2 C^* N |\mathcal{M}| +  \left(N |\mathcal{M}|\right)^2 - 4 N |\mathcal{M}| C^*\nonumber\\
    &=& \left(C^* - N |\mathcal{M}|\right)^2 \geq 0,
\end{IEEEeqnarray}
and $N |\mathcal{M}| C^* > 0$.
Hence, \eqref{eq:steady_state} has two real roots
\begin{align}
    i^{\infty}_{1,2} = \frac{1}{2} &\left\lbrace\left(1+\frac{a \kd}{\ka}\right)C^* + N |\mathcal{M}|\right.\nonumber\\
    & {} \left.\pm \sqrt{\left[\left(1+\frac{a \kd}{\ka}\right)C^* + N |\mathcal{M}|\right]^2 - 4 N |\mathcal{M}| C^*}\right\rbrace
\end{align}
and 
\begin{align}
    i^{\infty}_{1} &\geq \frac{1}{2} \left\lbrack\left(1+\frac{a \kd}{\ka}\right)C^* + N |\mathcal{M}| + |C^* - N |\mathcal{M}||\right\rbrack\nonumber\\
    &> \frac{1}{2} \left\lbrack C^* + N |\mathcal{M}| + |C^*| - |N |\mathcal{M}||\right\rbrack = C^*.
\end{align}

\noindent
Since values $i^{\infty} > C^*$ are infeasible, $i^{\infty}=i^{\infty}_2$.
This concludes the proof.

\subsection{Proof of Lemma~\ref{lem:steady_state_approximation}}\label{sec:app:proof_lemma_1}

According to Theorem~\ref{thm:steady_state}, $\iinfty(N)$ is defined as
\begin{align}
    i&=\iinfty(N)\nonumber\\
    &= \frac{1}{2} \left\lbrace\left(1+\lambda\right)C^* + N - \sqrt{\left[\left(1+\lambda\right)C^* + N \right]^2 - 4 N C^*}\right\rbrace,\label{eq:def_i_N}
\end{align}
where we have defined $\frac{a \kd}{\ka}=\lambda$ for ease of notation.
Introducing $\epsilon = j/N$, we write $\iinfty(N-j)$ as
\begin{multline}
    \tilde{i}(\epsilon) = \iinfty(N-j) = \frac{N}{2} \left\lbrace\left(1+\lambda\right)\frac{C^*}{N} + 1 - \epsilon\right.\\
    {}- \left.\sqrt{\left[\left(1+\lambda\right)\frac{C^*}{N} + 1 - \epsilon \right]^2 - 4 (1-\epsilon) \frac{C^*}{N}}\right\rbrace.
\end{multline}

\noindent
Next, we expand $\tilde{i}(\epsilon)$ in a Taylor series at $\epsilon=0$ and obtain
\begin{IEEEeqnarray}{rl}
    \IEEEeqnarraymulticol{2}{l}{
    \tilde{i}(\epsilon)}\nonumber\\
    =& \frac{1}{2}\left\lbrace\left(1+\lambda\right)C^* + N - \sqrt{\left[\left(1+\lambda\right)C^* + N \right]^2 - 4 N C^*}\right\rbrace\nonumber\\
    &{}-\frac{N}{2}\left\lbrace 1 - \frac{\left(1+\lambda\right)C^* + N - 2C^*}{\sqrt{\left[\left(1+\lambda\right)C^* + N \right]^2 - 4 N C^*}}\right\rbrace\epsilon\nonumber\\
    &{}-\frac{(N C^*)^2 \lambda}{\left\lbrace\left[\left(1+\lambda\right)C^* + N \right]^2 - 4 N C^*\right\rbrace^{3/2}}\epsilon^2\nonumber\\
    &{}-\frac{N \left(N + (1 + \lambda)C^* - 2 C^*\right)(N C^*)^2 \lambda}{\left\lbrace\left[\left(1+\lambda\right)C^* + N \right]^2 - 4 N C^*\right\rbrace^{5/2}}\epsilon^3 + O(\epsilon^4).\nonumber\\*\label{eq:steady_state_approx_expansion}
\end{IEEEeqnarray}

As we are interested in modeling the nonlinear impact of $N$ on $i$, we do not want to discard the higher-order terms in \eqref{eq:steady_state_approx_expansion}.
Instead, after repeatedly exploiting \eqref{eq:def_i_N} and \eqref{eq:steady_state}, we rewrite \eqref{eq:steady_state_approx_expansion} as
\begin{IEEEeqnarray}{rl}
    \IEEEeqnarraymulticol{2}{l}{
    \tilde{i}(\epsilon) = i - \frac{i \left(C^* - i\right)}{N C^*\left[ 1 - i^2/(N C^*) \right]}N \epsilon}\nonumber\\
    &{}-\frac{i^3 C^* \left(i - C^* - N + (N C^*)/i\right)}{(N C^*)^3 \left[ 1 - i^2/(N C^*) \right]^3}(N \epsilon)^2\nonumber\\
    &{}-\frac{\left[i + (N C^*)/i - 2 C^*\right] i^5 C^* \left(i - C^* - N + (N C^*)/i\right)}{(N C^*)^5 \left[ 1 - i^2/(N C^*) \right]^5} \nonumber\\
    &{}\times (N\epsilon)^3 + O(\epsilon^4).\label{eq:steady_state_expansion_simplified}
\end{IEEEeqnarray}
Now, we use our knowledge of the physical system to further simplify \eqref{eq:steady_state_approx_expansion}.
Namely, we know that only a small fraction of the solute molecules in our system are actually exposed to receptors, because they are spread all over the domain.
Furthermore, we have assumed $N \gg C^*$.
Hence, as $i \leq C^*$ by definition, we conclude that $i/N \ll 1$, $i^2/(N C^*) \ll 1$, and simplify \eqref{eq:steady_state_expansion_simplified} to
\begin{IEEEeqnarray}{rCl}
    \tilde{i}(\epsilon) &\approx& i - \frac{i \left(C^* - i\right)}{N C^*} N \epsilon - \frac{i^2 \left(C^*- i\right)}{(N C^*)^2}(N \epsilon)^2\nonumber\\
    &&{}- \frac{i^3 \left(C^*-i\right)}{(N C^*)^3}(N \epsilon)^3 + O(\epsilon^4)\nonumber\\
    &=&i \left[1 - \frac{C^* - i}{N C^*} N \epsilon - \frac{i \left(C^*- i\right)}{(N C^*)^2}(N \epsilon)^2\right.\nonumber\\
    &&{}- \left.\frac{i^2 \left(C^*-i\right)}{(N C^*)^3}(N \epsilon)^3 \right]+ O(\epsilon^4).\label{eq:steady_state_approximation_trunc_sum}
\end{IEEEeqnarray}
Finally, we use the identity $\sum_{k=1}^{\infty}z^k = z/(1-z)$ to rewrite \eqref{eq:steady_state_approximation_trunc_sum} as
\begin{align}
    \tilde{i}(\epsilon) &\approx i \left[1 - \frac{C^* - i}{i} \frac{(i N \epsilon)/(N C^*)}{1 - (i N \epsilon)/(N C^*)} \right]\nonumber\\
    &= i \left[ 1 - \frac{N \epsilon \left(C^* - i\right)}{N C^* - i N \epsilon}\right] = i \left[\frac{N C^* -N \epsilon C^*}{N C^* - i N \epsilon}\right]\nonumber\\
    &= i \left[\frac{N - N \epsilon}{N - N \epsilon (i/C^*)}\right].\label{eq:steady_state_approx_epsilon}
\end{align}

\noindent
This substitution is accurate to $O(\epsilon^4)$ as only terms up to order $4$ are considered in \eqref{eq:steady_state_approximation_trunc_sum}, while the series $\sum_{k=1}^{\infty}z^k$ involves higher-order terms.
Substituting $\epsilon = j/N$ in \eqref{eq:steady_state_approx_epsilon}, we obtain \eqref{eq:lemma_scaling_law}.
This concludes the proof.
\subsection{Proof of Theorem~\ref{thm:PI_PH}}\label{sec:app:proof_thm_2}
We start from \eqref{eq:statistical_model}:
\begin{IEEEeqnarray}{rCl}
    \IEEEeqnarraymulticol{3}{l}{P_{I}(n)}\nonumber\\
    &=& \binom{N}{n} P_{M_1,\ldots,M_n,M_{n+1},\ldots,M_N}(1,\ldots,1,0,\ldots,0)\nonumber\\
    &=&\binom{N}{n} P_{M_N|M_1,\ldots,M_n,M_{n+1},\ldots,M_{N-1}}(0|1,\ldots,1,0,\ldots,0)\nonumber\\
    &&{}\times\ldots\times P_{M_1}(1)\nonumber\\
    &\stackrel{\eqref{eq:statistical_model_steady_state}}{\approx}& \prod_{j=0}^{n-1}\frac{i}{j+1}\left( 1 - \frac{j}{C^*} \right) \frac{N-j}{N-j(i/C^*)}\nonumber\\
    &&{}\times\prod_{l=0}^{N-n-1}\left(1-\frac{i}{N-n-l}\left( 1 - \frac{n}{C^*} \right)\right.\nonumber\\
    &&\qquad\qquad\left.\times \frac{N-n-l}{N-(n+l)(i/C^*)}\right)\nonumber\\
    &=& \frac{N!}{n!(N-n)!} \prod_{j=0}^{n-1}\left( i - \frac{j i}{C^*} \right) \frac{1}{N-j(i/C^*)}\nonumber\\
    &&{}\times\prod_{l=0}^{N-n-1}\left(1-i\left( 1 - \frac{n}{C^*} \right) \frac{1}{N-(n+l)(i/C^*)}\right)\nonumber\\
    &=&\binom{N}{n}\prod_{j=0}^{n-1}\frac{C^* - j}{NC^*/i-j}\prod_{l=0}^{N-n-1}\frac{NC^*/i - C^* - l}{NC^*/i-n-l}\nonumber\\
    &=&\frac{\binom{C^*}{n}\binom{NC^*/i - C^*}{N - n}}{\binom{NC^*/i}{N}} = P_{H}(n;NC^*/i,C^*,N).\label{eq:thm_hypergeometric_proof}
\end{IEEEeqnarray}

\noindent
We note that since $C^*/i \geq 1$ and, by assumption, $i \leq \frac{C^*}{1+C^*/N}$, \eqref{eq:thm_hypergeometric_proof} is always valid for $n \leq C^*$.
Finally, we note that $i \leq \frac{C^*}{1+C^*/N}$ is a mild technical assumption which is justified by the observation that $i$ can only be close to $C^*$ if $N$ is large.
However, if $N$ is large, $i \leq \frac{C^*}{1+C^*/N}$ reduces to $C^*/i \geq 1$ which is fulfilled by definition.
This completes the proof.
    
\bibliographystyle{IEEEtran}    
\bibliography{IEEEabrv,sl_bib,%
    icc21_max_bib%
}
\end{document}